\documentclass[10pt,a4paper]{article}
\usepackage[a4paper,left=2.5cm,right=2.5cm,top=2.5cm,bottom=2.5cm]{geometry}
\usepackage[super,compress]{cite}
\usepackage{graphicx}
\usepackage{authblk}
\usepackage{xcolor}

\begin{document}

\title{Constant-roll inflation with tachyon field in the holographic braneworld}


\author[1]{Marko Stojanovic\thanks{marko.stojanovic@pmf.edu.rs}}
\author[2]{Neven Bili\'c\thanks{bilic@irb.hr}}
\author[3]{Dragoljub D. Dimitrijevic\thanks{ddrag@pmf.ni.ac.rs}}
\author[3]{Goran S. Djordjevic\thanks{gorandj@junis.ni.ac.rs}}
\author[3]{Milan Milosevic\thanks{milan.milosevic@pmf.edu.rs}}

\affil[1]{Faculty of Medicine, University of Ni\v s, Serbia}
\affil[2]{Division of Theoretical Physics, Rudjer Bo\v{s}kovi\'{c} Institute, Zagreb, Croatia}
\affil[3]{Department of Physics, University of Ni\v s, Serbia}

\maketitle

\begin{abstract}
We study inflation driven by the tachyon field in the holographic braneworld by assuming the second slow-roll parameter $\eta$ is constant. The parameter $\eta$ can be either defined by the tachyon scalar field and the Hubble parameter or by the Hubble parameter only. By assuming a constant $\eta$, we derive and numerically solve a differential equation for the Hubble expansion rate. We calculate numerically the scalar spectral index and the tensor-to-scalar ratio. We confront the results with the observational data and find some constraints on the free model parameters. The swampland conjectures are discussed in the context of the constant-roll inflation, with some accent on the holographic model. 
\end{abstract}


\section{Introduction}  

Inflation is the most promising candidate for solving the horizon problem and other related problems of the standard Big Bang cosmology [\citen{Guth:1980zm}], by demanding an extremely short period in the early universe during which the universe rapidly expands. Moreover, the quantum fluctuations generated during inflation lead to the primordial density perturbation responsible for the large-scale structure formation.

An outstanding achievement has been made in  the theoretical prediction of the inflation parameters, assuming that the inflation  is governed  by a single scalar field (the inflaton), whose origin is still unknown. One of the popular classes of inflationary models is based on the dynamics of the tachyon field originating in M or string theory [\citen{Sen:2002nu}]. It is known [\citen{Fairbairn:2002yp}] that the decay of unstable D-branes in string theory is followed by a production of pressureless gas with non-zero energy density, which is required for a phase of inflationary expansion. This is one of main motivations for applications of the tachyon field in the theory of inflation.

The constant-roll inflation is based on the assumption that during inflation, the second slow-roll parameter $\eta$ is an arbitrary constant and not necessarily small [\citen{Motohashi:2014ppa}]. This formalism has been applied in many different scenarios of inflation [\citen{Gao:2018tdb,Mohammadi:2018oku,Anguelova:2017djf,Yi:2017mxs,Odintsov:2019ahz,Lahiri:2018erp,Mohammadi:2020ftb,Nojiri:2017qvx,Nguyen:2021emx}]. In this work, for the first time, we apply the formalism of the constant-roll inflation to the holographic second Randall-Sundrum (RSII) model with a tachyon field [\citen{Bilic:2018uqx,Milosevic:2018gck}]. This paper is an extension of our previous work in which we studied the constant-roll tachyon inflation in the  RSII cosmology [\citen{Stojanovic:2023qgm}]. We have shown that the observational parameters obtained in the constant-roll approach for the RSII model are in better agreement with observational data than those obtained in the slow-roll approximation [\citen{Bilic:2016fgp}]. 

Motivated by this result, we examine whether similar behavior will occur in the tachyon model in the holographic braneworld. The dynamics of the model considered here are based on a scenario in which the brane, with an effective tachyon field, is located at the boundary of a 5-dim asymptotically Anti-de Sitter (AdS\textsubscript{5}) space. This type of model we refer to as the holographic braneworld [\citen{Bilic:2015uol}]. The effective four-dimensional Einstein equations on the holographic boundary of AdS\textsubscript{5} yield a modified Friedmann equation with a quartic term $\propto H^{4}$, where $H$ is Hubble parameter [\citen{Apostolopoulos:2008ru,Bilic:2015uol}]. Due to this modification in the Friedmann equation, it is interesting to consider the constant-roll inflation in the (holographic) cosmology on the AdS\textsubscript{5} boundary. The constant-roll inflation has recently been analyzed, assuming that inflation  is driven by (generalized) holographic dark energy [\citen{Nojiri:2023nop,Mohammadi:2022vru}]. It is important to stress that these models are not equivalent to our model.

The paper is organized as follows. In Section 2,  we give a brief review of the tachyon dynamics in the holographic braneworld. The scenario of the constant-roll inflation in the holographic model is discussed in Section 3. In Section 4, we briefly review the calculation of the primordial power spectra and present our numerical results. In Section 5, we discuss the swampland criteria for the holographic constant-roll RSII inflation model. A summary and conclusions are given in Section 6.

\section{Tachyon cosmology in the holographic braneworld}
We will consider two types of braneworlds in the AdS$_5$  bulk: the second Randall-Sundrum (RSII) model and the holographic braneworld.
	In the RSII model [\citen{randall2}], a  3-brane is
	located at a finite distance from the boundary of AdS$_5$.
	The model was originally proposed as an alternative to
	the compactification of extra dimensions.
	In a holographic braneworld, a 3-brane is located 
	at the boundary of the asymptotic AdS$_5$ and 
	the effective Einstein equations on the brane are derived in the framework of anti-de Sitter/conformal field theory (AdS/CFT) correspondence and holographic renormalization [\citen{haro}].

\subsection{Randall-Sundrum and and holographic cosmologies}
\label{connection}

The first Randall-Sundrum model, usually referred to as the RSI model  [\citen{randall1}], is proposed as a solution to the hierarchy problem. The RSI model  consists of two branes of opposite tensions
embedded in the AdS$_5$ bulk,
where the observer brane is the brane with negative tension.
In the RSII model, the observers are placed on a single positive tension brane.
In a holographic braneworld universe, a 3-brane is located 
at the boundary of the asymptotic AdS$_5$.  In this way, the holographic cosmology may be regarded as a limiting case of the
RSII cosmology. The braneworld univerrse evolution is
governed by matter on the brane in addition to the
CFT fluid.

A cosmology on the brane is obtained by allowing the brane to
move in the bulk along the fifth dimension $z$. Equivalently, the brane is kept fixed at $z = z_{\rm br}$
while making the metric in the bulk time dependent.
The time-dependent bulk spacetime with line element  
\begin{equation}
	ds_{(5)}^2=\frac{\ell^2}{z^2}\left( 
	n^2(\tau,z)d\tau^2- a^2(\tau,z) d\Omega_\kappa^2-dz^2 
	\right),
	\label{eq102}
\end{equation}
may be regarded as
a $z$ foliation of the bulk with an FRW cosmology on each $z$ slice.
In particular, 
at $z=z_{\rm br}$ we have the RSII cosmology and, at z=0, the holographic cosmology.
The Friedmann equation on the RSII brane is modified
\begin{equation}
	H_{\rm br}^2+\frac{\kappa}{a_{\rm br}^2}=\frac{8\pi G_{\rm N}}{3} \rho_{\rm br} +
	\left(\frac{4\pi G_{\rm N}\ell}{3}\right)\rho_{\rm br}^2
	+\frac{\mu\ell^2}{a_{\rm br}^4} ,
	\label{eq022}
\end{equation}
where $\ell$ is the AdS$_5$ curvature radius, $H_{\rm br}=\dot{a}_{\rm br}/(n_{\rm br}a_{\rm br})$ is the Hubble rate on the RSII brane.
Equation
(\ref{eq022}) was derived assuming the fine-tuned value 
of the brane tension $\lambda=3m_{\rm Pl}^2/(4\pi\ell^2)$ 
which assures the vanishing of the cosmological constant.
In the last term on the right-hand side of (\ref{eq022}), usually referred to as the "dark radiation", the parameter
$\mu$ is related to the bulk black-hole mass
$\mu=(8G_5 M_{\rm bh})/(3\pi \ell^2$  (see, e.g, [\citen{maartens}]).
	
By introducing the boundary in AdS$_5$ at
$z = z_{\rm br}$ instead of $z = 0$, the RSII model is conjectured to be dual to
a cutoff CFT coupled to gravity,
with $z=z_{\rm br}$ providing
the IR cutoff. 
This conjecture  then reduces to
the standard AdS/CFT duality as the boundary is pushed off to $z=0$. 
In this limit, the variation of the action yields Einstein's equations on the
boundary [\citen{Bilic:2015uol,Apostolopoulos:2008ru}]
\begin{equation}
	R_{\mu\nu}- \frac12 R g_{\mu\nu}= 8\pi G_{\rm N} (\langle T^{\rm CFT}_{\mu\nu}\rangle +T^{\rm matt}_{\mu\nu}),
	\label{eq3006}
\end{equation}
where 
$T^{\rm matt}_{\mu\nu}$ is the energy-momentum tensor associated with matter on the holographic brane 
and  $T^{\rm CFT}_{\mu\nu}$ the energy-momentum tensor of
the CFT on the boundary.
According to the AdS/CFT prescription, the expectation value $\langle T^{\rm CFT}_{\mu\nu}\rangle$
is obtained 
by functionally differentiating the renormalized on-shell bulk gravitational action with respect to the
boundary metric [\citen{haro}]. 
From (\ref{eq3006}), one derives the Friedmann equation at the boundary
\begin{equation}
	\mathcal{H}^2=\frac{\ell^2}{4}
	\left(\mathcal{H}^4+ \frac{4\mu}{a^4} \right)+\frac{8\pi G_{\rm N}}{3}\rho.
	\label{eq3110}
\end{equation}
where $\mathcal{H}^2=H^2+\kappa/a^2$ and $H=\dot{a}/a$ is the Hubble rate
at the holographic boundary. As in the RSII model, the parameter
$\mu$ is related to the black hole (BH) in the bulk, and the term proportional to $\mu$  is dubbed the "dark radiation".
The second Friedmann equation may be obtained by combining the time derivative of (\ref{eq3110})
with the energy conservation equation
\begin{equation}
	\partial_t\rho+3(\rho+p)\frac{\partial_t a}{a}=0.
	\label{3201}
\end{equation}	
One finds
\begin{equation}
	\frac{\ddot{a}}{a}
	\left(1-\frac{\ell^2}{2}\mathcal{H}^2\right)+\mathcal{H}^2=
	\frac{4\pi G_{\rm N}}{3}(\rho-3p).
	\label{eq3113}
\end{equation}

	A map between RSII  and holographic cosmologies  can
	be constructed using [\citen{Apostolopoulos:2008ru}]
	\begin{equation}
		a_{\rm br}^2=a^2\left[
		\left(1-\frac{\mathcal{H}^2 z^2}{4}\right)^2
		+ \frac14 \frac{\mu z^4}{a^4}
		\right],
		\quad
		n_{\rm br}=\frac{\dot{a}_{\rm br}}{\dot{a}}.
		\label{eq3103}
	\end{equation}
	The Hubble rates are related by
	\begin{equation}
		\mathcal{H_{\rm br}}\equiv H_{\rm br}^2+\frac{\kappa}{a_{\rm br}^2} = \mathcal{H}\frac{a}{a_{\rm br}}.
		\label{eq201}
	\end{equation}
	Using this and (\ref{eq3103}) we can find a relation between the cosmological scales $a_{\rm br}$  on the
	brane at $z=z_{\rm br}$ and $a_z$ at on an arbitrary $z$-slice.
	First, we can express the first equation in (\ref{eq3103}) as an equation for $a^2$, $a_z^2$, and $\mathcal{H}_z^2$, 
	and similarly as another equation for $a^2$, $a_{\rm br}$, and $\mathcal{H}_{\rm br}^2$ at $z=z_{\rm br}$.
	By eliminating $a^2$ from  these
	two equations  we find a relation between the scale $a_z$ and $a_{\rm br}$.
Then, using this relation and   (\ref{eq201}) we can express the Hubble rate $H$ at $z=0$ in terms of
	the Hubble rate $H_{\rm br}$ at $z=z_{\rm br}$
	\begin{equation}
		\mathcal{H}^2= 2\mathcal{H}_{\rm br}^2
		\left(
		1 +\frac{\mathcal{H}_{\rm br}^2 z_{\rm br}^2}{2} \pm
		\sqrt{1+\mathcal{H}_{\rm br}^2 z_{\rm br}^2
			-\frac{\mu z_{\rm br}^4}{a_{\rm br}^4}}
		\right)^{-1} .
		\label{eq3209}
	\end{equation}
This equation provides a direct link between RSII and holographic cosmologies.

\subsection{The tachyon dynamics in the holographic braneworld}

In this section, we briefly introduce the basic property of the tachyon dynamics in the holographic braneworld.   For a spatially flat FLRW geometry on the holographic brane, the modified  Friedmann equations (\ref{eq3110}) and (\ref{eq3113})
can be written as
\begin{equation}
	h^2-\frac{1}{4}h^4=\frac{\kappa^2}{3}\ell^4\rho,\label{F1}
\end{equation}
\begin{equation}
	\dot{h}\left(1-\frac{1}{2}h^2\right)=-\frac{\kappa^2}{2} \ell^3(p+\rho),
	\label{F2}
\end{equation}
where $p$ and $\rho$ are the pressure and the density of the fluid associated with the field on the brane and 
\begin{equation}
	h\equiv \ell H.
	\label{eq2015}
\end{equation}
is a dimensionless Hubble expansion rate. The overdot in (\ref{F2}) denotes a derivative with respect to time   measured in units of $\ell$. The fundamental coupling parameter $\kappa$ is related to the AdS\textsubscript{5} curvature radius $\ell$  and the four-dimensional Newton constant $G_{\rm N}$
\begin{equation}
	\kappa^{2}=\frac{8\pi G_{\rm N}}{\ell^2}.
\end{equation}

As mentioned previously, Eqs.\
 (\ref{F1}) and (\ref{F2}) might contain 
the term responsible for the so called "dark radiation" and the cosmological constant term [\citen{Bilic:2015uol,Bilic:2018cyh}]. 
We have omitted these terms for the following reasons. First, as in RSII cosmology,  
dark radiation is related to the BH in the AdS bulk.  Since dark radiation is not relevant for inflation, we  
assume that the bulk is pure AdS$_5$ with no BH, and hence, the terms responsible for it do not appear in our  modified Friedmann equations  (\ref{F1}) and (\ref{F2}).
Second,  by invoking the so-called RSII fine-tuning condition, we have adjusted the brane tension so that the cosmological constant vanishes. In principle,
as discussed in Ref.\ [\citen{Bilic:2015uol}],  one could fine-tune the brane tension so that 
a tiny nonvanishing cosmological constant remains. This type of fine tuning may be relevant for recent cosmology. However, 
during inflation, the contribution of the cosmological constant is negligible   since the tachyon field causes a fast quasi-de Sitter expansion anyway.

The physically acceptable solution to the equation (\ref{F1}), as a quadratic equation for $h^{2}$, is of the form
\begin{equation}
	h^{2}=2\left(1-\sqrt{1-\frac{\kappa^2}{3}\ell^4\rho}\right).
	\label{F1sol}
\end{equation}
In the low-density limit ($\kappa^{2}\ell^{4}\rho\ll 1$), this equation reduces to the standard Friedmann equation, to wit, 
the holographic cosmology reduces to the standard cosmology.

As discussed in Section \ref{connection} (for more details, see Ref.\ [\citen{Bilic:2015uol}]), the holographic model is a limiting case of the RSII model with a single positive tension brane
approaching the boundary of AdS$_5$. Inflation based on the RSII model
can have arbitrary large (up to $m_{\rm Pl}$) Hubble scale,  
depending on the ratio $\rho/\lambda$, where $\lambda$ is the brane tension (see, e.g., Ref.\ [\citen{maartens}]).
For exaample, if $\rho\sim \lambda$, one finds $H \sim 3/\ell$. In contrast, our Hubble scale is constrained by
Eq.\ (\ref{F1sol}) and cannot exceed  $H_{\rm max}\equiv\sqrt{2}/\ell$.
According to our estimate (Section  \ref{estimate} further ahead), $\ell \sim 4\times 10^5 l_{\rm Pl} $, 
and hence, 
the maximal Hubble scale in our holographic model $H_{\rm max}\sim 10^{-6}m_{\rm Pl}$ is way below the Planck scale.

According to (\ref{F2}), 
 the expansion rate is a monotonously decreasing function of time. 
 In this model, the evolution is constrained by the physical range of the expansion rate $0\leq h^{2}\leq 2$. 
 The evolution of the universe starts with an initial $h_{\rm i}<\sqrt{2}$ and with negative $\dot{h}_{\rm i}$.

\subsection{Tachyon fluid}
 
For a fluid on the brane described by the tachyon field, the tachyon Lagrangian of the Dirac-Born-Infeld (DBI) type is of the form 
\begin{equation}
{\cal{L}} = -\ell^{-4} V(\theta/\ell)
\sqrt{1-g^{\mu\nu}\theta_{,\mu}\theta_{,\nu}},
\end{equation}
where the tachyon field $\theta$ has units of $\ell$. The tachyonic potential $V(\theta/\ell)$ satisfies the following properties
\begin{equation}
V(0)={\rm const},\quad V_{,\theta}(\theta>0)<0,\quad V(|\theta|\rightarrow\infty)\rightarrow 0,
\label{vprop}
\end{equation}
where the subscript $,\theta$ denotes a derivative with respect to $\theta$. The pressure $p$ and energy density $\rho$ are the components of the energy-momentum tensor $T_{\mu\nu}$
\begin{equation}
T_{\mu\nu}=(\rho+p)u_{\mu}u_{\nu}-pg_{\mu\nu},
\end{equation}
where the components of the four velocity are $u_{\mu}\equiv \partial_{\mu}\theta/\sqrt{\partial_{\mu}\theta\partial^{\mu}\theta}$ and $u_{\alpha}u^{\alpha}=1$. Assuming spatial isotropy and homogeneity at the background, $\theta=\theta(t)$, from 
\begin{equation}
T_{\mu\nu}\equiv \frac{2}{\sqrt{-g}}\frac{\delta \sqrt{-g}{\cal{L}}}{\delta g^{\mu\nu}}, 
\end{equation}
one finds 
\begin{equation}
p=-\ell^{-4}V\sqrt{1-\dot{\theta}^2},
\label{p}
\end{equation}
\begin{equation}
\rho=\frac{\ell^{-4}V}{\sqrt{1-\dot{\theta}^2}}.
\label{rho}  
\end{equation}
The model is presented in more detail in Ref.\ [\citen{Bilic:2018uqx}]. In the following, we will analyze the constant-roll inflation based on this model, with the slow-roll parameter $\eta$ being constant.

\section{Constant-roll inflation}

The dominant inflationary scenario nowadays, known as the slow-roll inflation, is based on the assumption that the inflaton field varies very slowly and inflation lasts long enough to overcome some problems in the standard Bing Bang cosmology.
In the slow-roll inflation models, the slow-roll parameters
\begin{equation}
\epsilon=-\frac{\ell\dot{h}}{h^2},
\label{defepsilon}
\end{equation}
\begin{equation}
\eta=\frac{\ell\ddot{\theta}}{h\dot{\theta}},
\label{defetanew}
\end{equation}
are small. Inflation ends when the parameter $\epsilon$ crosses unity. The modification of the slow-roll inflation ($\eta\approx 0$) or ultra slow-roll inflation ($\eta=-3$) leads to the constant-roll inflation defined by the condition [\citen{Motohashi:2014ppa}] 
\begin{equation}
\eta=\rm{const}.
\label{eta}
\end{equation}
In this work, we apply the condition (\ref{eta}) to the tachyon dynamics in the holographic braneworld. Combining Friedmann’s equations, (\ref{F1}) and (\ref{F2}), with (\ref{p}) and (\ref{rho}) we obtain 
\begin{equation}
\dot{\theta}=-\frac{2\ell}{3}\frac{h,_{\theta}}{h^2}\frac{1-\frac{1}{2}h^2}{1-\frac{1}{4}h^2},
\label{dottheta}
\end{equation}
where according to the Hamilton-Jacobi formalism the expansion rate $h$  is taken as a function of field $\theta$, i.e. $h=h(\theta)$, yielding $\dot{h}=h_{,\theta}\dot{\theta}$. 
By taking the time derivative of (\ref{dottheta}) we obtain
\begin{equation}
\ddot{\theta}=-\frac{2\ell}{3}\frac{1-\frac{1}{2}h^2}{1-\frac{1}{4}h^2}\frac{1}{h^3}\left(hh_{,\theta\theta}-2h_{,\theta}^2\right)\dot{\theta}
+\frac{\ell}{3}\frac{h_{,\theta}^2}{h}\frac{1}{(1-\frac{1}{4}h^2)^2}\dot{\theta}.
\label{ddottheta}
\end{equation}
Then, using the condition (\ref{eta}), the equations (\ref{dottheta}) and (\ref{ddottheta}), and replacing them in the equation (\ref{defetanew}) we derive the differential equation for the expansion rate 
\begin{equation}
hh_{,\theta\theta}-2h,_{\theta}^2\left(1+\frac{h^2}{4(1-\frac{1}{2}h^2)(1-\frac{1}{4}h^2)}\right)+\frac{3}{2\ell^2}\eta h^4\frac{1-\frac{1}{4}h^2}{1-\frac{1}{2}h^2}=0.\label{hthetatheta}
\end{equation}
The equation (\ref{hthetatheta}) is more complicated than the corresponding equation in the tachyon constant-roll model in standard cosmology (derived from the Einstein-Hilbert action) and will also be solved numerically. 

One of the main advantages of constant-roll inflation is that there is no need to specify the type of potential. The potential can be reconstructed when the dynamics of the model is known. Combining (\ref{F1}), (\ref{rho}) and (\ref{dottheta}), we obtain the potential as an implicit function of the tachyon field 
\begin{equation}
V=\frac{3}{\kappa^2}h^2(1-\frac{1}{4}h^2)\sqrt{1-\frac{4\ell^2}{9}\frac{h_{,\theta}^2}{h^4}\left(\frac{1-\frac{1}{2}h^2}{1-\frac{1}{4}h^2}\right)^2}.\label{potential}
\end{equation}
In the limit when $h^{2}\ll 1$, the equations (\ref{hthetatheta}) and (\ref{potential}) agree with those derived in tachyon constant-roll model in standard cosmology [\citen{Mohammadi:2018oku}].

There are plenty of definitions of the slow-roll parameters. Besides the most common, given by (\ref{defepsilon}) and (\ref{defetanew}), it proves advantageous to define the slow-roll parameters recursively
\begin{equation}
\varepsilon_{0}\equiv h_{*}/h,
\end{equation}
\begin{equation}
\varepsilon_{i+1}\equiv \frac{d\ln|\varepsilon_{i}|}{dN},\quad i\ge 0,
\end{equation}
where $h_{*}$ is the expansion rate at an arbitrarily chosen time and $N$ is the number of e-folds at time $t$ defined as
\begin{equation}
N(t)=\frac{1}{\ell}\int_{t_{\rm i}}^t h dt,
\label{Ndef}
\end{equation}
where $t_{\rm i}$  denotes the time at the beginning of inflation.
Using (\ref{dottheta}), Eq.\ (\ref{Ndef}) may be expressed as
\begin{equation}
	N(\theta)=\frac{1}{\ell}\int_{\theta_{\rm i}}^\theta \frac{h}{\dot{\theta}}d\theta=-\frac{3}{2\ell^2}\int_{\theta_{\rm i}}^\theta \frac{h^3}{h,_{\theta}}\frac{(1-\frac{1}{4}h^2)}{(1-\frac{1}{2}h^2)}d\theta,
	\label{efolds}
\end{equation}
 To solve the most important problems in the early universe, the number of e-folds at the end of inflation should be $N\equiv N(\theta_{\rm f})\simeq 60$, 
 where $\theta_{\rm f}=\theta(t_{\rm f})$ is fixed by $\varepsilon_{1}(\theta_{\rm f})=1$. 
 
 From the definition for the parameters $\varepsilon_{i}$ one obtains  
\begin{equation}
\varepsilon_{i+1}=\frac{\ell\dot{\varepsilon}_{i}}{h\varepsilon_{i}}.
\label{Ndef1}
\end{equation}
The relationships between $\varepsilon_{1}\equiv \epsilon$, $\varepsilon_{2}$ and $\eta$ depend on the form of the Friedmann equation.
Using (\ref{Ndef1}) together with (\ref{dottheta}), for the first two slow-roll parameters $\varepsilon_{i}$ we obtain
\begin{equation}
\varepsilon_{1}
=-\ell\frac{\dot{h}}{h^2}
=\frac{2\ell^2}{3}\frac{h_{,\theta}^2}{h^4}\frac{1-\frac{1}{2}h^2}{1-\frac{1}{4}h^2},
\label{e1}
\end{equation}
\begin{equation}
\varepsilon_{2}
=\ell\frac{\dot{\varepsilon}_{1}}{h\varepsilon_{1}}
=\eta-\varepsilon_{1}\left(\frac{h_{,\theta\theta}}{h_{,\theta}^2}h-2\right).
\label{e21}
\end{equation}
Substituting $h_{,\theta\theta}$ from (\ref{hthetatheta}) into (\ref{e21}), $\varepsilon_{2}$ takes the following simple form
\begin{equation}
\varepsilon_{2}=2\eta-\frac{h^2}{2(1-\frac{1}{2}h^2)(1-\frac{1}{4}h^2)}\varepsilon_{1}.
\label{e22}
\end{equation}
In our model, unlike the constant-roll inflation model in the standard cosmology [\citen{Steer:2003yu}], the expression for $\varepsilon_{2}$ contains the additional term proportional to $h^{2}$, causing the change of $\varepsilon_{2}$ during the time. The parameter $\varepsilon_{3}$ is equal to 
\begin{equation}
\varepsilon_{3}=\frac{1}{\varepsilon_{2}}\left(\frac{h^2(1-\frac{1}{8}h^4)}{(1-\frac{1}{2}h^2)^2(1-\frac{1}{4}h^2)^2}\varepsilon_{1}^2 -\frac{h^2}{2(1-\frac{1}{2}h^2)(1-\frac{1}{4}h^2)}\varepsilon_{1}\varepsilon_{2}\right).
\label{e33}
\end{equation}
In the limit when $h^{2}\ll 1$, the contribution of the additional term can be neglected, i.e., $\varepsilon_{2}\simeq 2\eta$, causes that the parameter $\varepsilon_{2}$ is almost constant and the parameter $\varepsilon_{3}$ tends to zero.

\section{Primordial power spectra and observational parameters}

Following  Ref.\ [\citen{Bertini:2020dvv}],
we first briefly review the calculations of the primordial power spectra
for a  general k-essence inflation in the holographic braneworld.

\subsection{Scalar perturbations}

Assuming a spatially flat background,  we introduce
the perturbed line element in the Newtonian gauge
\begin{equation}
	ds^2=(1+2\Psi) dt^2-(1-2\Phi)a^2(t)(dr^2+r^2 d\Omega^2) . \label{eq0013}
\end{equation}
Inserting the above metric components in the field equations, we obtain
the Einstein equations at linear order. As shown in  Ref.\ [\citen{Bertini:2020dvv}], 
one finds $\Phi/\Psi=1$ as in general relativity, and hence,  
one can work in longitudinal gauge with $\Psi=\Phi$. 
Then, the relevant Einstein equations at linear order are given by
\begin{equation}
	a^{-2} \nabla^2 \Phi -3H\dot{\Phi} +3H^2\Phi=4\pi G_{\rm N}
	\delta {T}^0_0 (1-h^2/2)^{-1},
	\label{eq0014}
\end{equation}
\begin{equation}
	(\dot{\Phi}+H\Phi)_{,i}=4\pi G_{\rm N}
	\delta {T}^0_i (1-h^2/2)^{-1}.
	\label{eq0015}
\end{equation}
The  perturbations of the stress tensor components $\delta {T}^\mu_\nu$ are obtained by employing the perfect fluid description and
the energy-momentum conservation. 
One finds [\citen{garriga}]
\begin{equation}
	\delta {T}^0_0=\frac{{p}+{\rho}}{{c}_{\rm s}^2}
	\left[\left(\frac{\delta\theta}{\dot{\theta}}\right)^.-\Phi\right]
	-3H({p}+{\rho}) \frac{\delta\theta}{\dot{\theta}}.
	\label{eq0041}
\end{equation}
\begin{equation}
	\delta {T}^0_i=({p}+{\rho})\left(\frac{\delta\theta}{\dot{\theta}}\right)_{,i} \, .
	\label{eq0043}
\end{equation}
As in Ref.\ [\citen{garriga}], we introduce  new functions
\begin{equation}
	\xi=\frac{a\Phi}{4\pi G_{\rm N}H}, \quad
	\zeta = \Phi+H\frac{\delta\theta}{\dot{\theta}}.
	\label{eq0021}
\end{equation}
The quantity $\zeta$ is gauge invariant and  measures the spatial curvature of comoving
(or constant-$\theta$) hyper-surfaces. Then,
using (\ref{eq0041}) and (\ref{eq0043}), equations (\ref{eq0014}) and (\ref{eq0015})
take the form
\begin{equation}
	a\dot{\xi}=z^2c_{\rm s}^2\zeta,
	\label{eq0016}
\end{equation}
\begin{equation}
	a\dot{\zeta}=z^{-2}\nabla^2\xi .
	\label{eq0017}
\end{equation}
Here
\begin{equation}
	z=\frac{a(\tilde{p}+\tilde{\rho})^{1/2}}{c_{\rm s}H} =
	\frac{a}{c_{\rm s}}\sqrt{\frac{\varepsilon_1}{4\pi G_{\rm N}}} ,
	\label{eq0029}
\end{equation}
where we have defined 
\begin{equation}
	\tilde{p}+\tilde{\rho}=(p+\rho)(1-h^2/2)^{-1}.
	\label{eq1034}
\end{equation}

By introducing the conformal time $\tau=\int dt/a$ and a new variable $v=z\zeta$, it is straightforward
to show  that  $v$ satisfies a second-order differential equation
\begin{equation}
	v''-c_{\rm s}^2 \nabla^2 v-\frac{z''}{z}v =0,
	\label{eq0022}
\end{equation}
where the primes denote derivatives with respect to $\tau$.
By making use of the Fourier transformation,
we also obtain the  mode-function equation
\begin{equation}
	v_q''+\left(c_{\rm s}^2q^2  -\frac{z''}{z}\right)v_q =0.
	\label{eq0032}
\end{equation}
At linear order in $\varepsilon_i$ in the slow-roll regime,  one  can use  the relations
\begin{equation}
	\tau=- \frac{1+\varepsilon_1}{aH}+\mathcal{O}(\varepsilon_1^2),
\quad  \frac{z''}{z}=\frac{\nu^2-1/4}{\tau^2},
	\label{eq0047}
\end{equation}
where
\begin{equation}
	\nu^2=\frac94+
	3\varepsilon_1+\frac32\varepsilon_2  .
	\label{eq0048}
\end{equation}
 We normalize the
solution to (\ref{eq0032}) so that  it satisfies the positive frequency asymptotic limit
\begin{equation}
	\lim_{\tau\rightarrow -\infty}v_q=\frac{e^{-ic_{\rm s}q\tau}}{\sqrt{2c_{\rm s}q}}.
	\label{eq0027}.
\end{equation}
Then the solution which up to a phase agrees with (\ref{eq0027})
is given by
\begin{equation}
	v_q=\frac{\sqrt{\pi}}{2}(-\tau)^{1/2} H_\nu^{(1)}(-c_{\rm s}q\tau),
	\label{eq0049}
\end{equation}
where $H_\nu^{(1)}$ is the Hankel function of the first kind.
This solution interpolates between subhorizon and superhorizon regimes, to wit, between short and long wavelength limits.

For a given wavenumber $q$, the potential term $z''/z$ in (\ref{eq0032}) can be neglected at sufficiently early times ($\tau \rightarrow - \infty$ )
when the physical wavelength of the perturbation $a/q$ is much smaller than
the sound  horizon $c_{\rm s}/H$. Hence, in this short wavelength limit,  the normalized solutions are of the form  (\ref{eq0027}).

The potential term $z''/z$ starts to dominate after these modes cross the
sound horizon. Then, the solution to (\ref{eq0032}) evolves into the long wavelength
solution 
\begin{equation}
v_q \simeq C_q z, 
\end{equation}
where the constant $C_q$ is fixed by matching both solutions in
the standard way. One finds 
\begin{equation}
	|C_q|= \frac{1}{\bar{z}\sqrt{2\bar{c}_{\rm s} q} }, 
\end{equation}
where $\bar{z}$ and $\bar{c}_{\rm s}$ are the values of $z$ and $c_{\rm s}$
at the moment of sound horizon crossing $aH=c_{\rm s}q$.
Hence,  the curvature perturbation $\zeta_q=v_q/z\simeq C_q$ is conserved 
in the superhorizon region.

The calculation of the power spectrum proceeds as usual.
 Applying the standard canonical quantization [\citen{mukhanov}],  the field $v_q$ is promoted to an operator
and the power spectrum of the field $\zeta_q$ is obtained  from the two-point correlation function.
The 
spectral density
\begin{equation}
	\mathcal{P}_{\rm S}(q)=\frac{q^3}{2\pi^2}|\zeta_q|^2=\frac{q^3}{2\pi^2z^2}|v_q|^2 ,
	\label{eq0024}
\end{equation}
with $v_q$ given by (\ref{eq0049}),
characterizes the primordial scalar fluctuations.
Evaluating this at the horizon crossing 
one finds
at the lowest order in $\varepsilon_1$ and $\varepsilon_2$
\begin{equation}
	\mathcal{P}_{{\rm S}} \simeq \frac{G_{\rm N} H^2 }{\pi c_{\rm s}\varepsilon_1 }
	\left[1-2\left(1+C\right)\varepsilon_1-C \varepsilon_2\right],
	\label{eq3007}
\end{equation}
where $C=\gamma-2+\ln 2 \simeq -0.73$ and $\gamma$ is the Euler constant,
so we recover the standard expression [\citen{Steer:2003yu,hwang}]. 

\subsection{Tensor perturbations}
\label{tensor}

The tensor perturbations are related to the production of gravitational waves
during inflation.
The metric perturbation is, in this case, written as
\begin{equation}
	ds^2= dt^2-a^2(t)\left(\delta_{ij}+h_{ij}\right)dx^idx^j,
	\label{eq0054}
\end{equation}
where $h_{ij}$ is traceless and transverse.
Inserting the metric components in the holographic field equations (see  Ref.\ [\citen{Bertini:2020dvv}] for details) we obtain
\begin{equation}
	\left(1-\frac{\ell^2}{2}(H^2+\dot{H})\right)
	\left(\ddot{h}_{ij}+3H \dot{h}_{ij}-\frac{1}{a^2}\nabla^2 h_{ij}\right)=8\pi G_{\rm N} \delta T_{ij}.
	\label{eq2055}
\end{equation}
In the absence of anisotropic stress, the gravitational waves decouple from matter, and the right-hand side of 
the above equation is zero. Then,  the expression in the first bracket on the left-hand side 
factorizes out, and hence, our braneworld scenario 
does not introduce changes to the gravitational wave dynamics.
At the lowest order in slow-roll parameters, the 
spectral density that characterizes the primordial tensor fluctuation
is given by the usual expression  [\citen{lidsey1}]
\begin{equation}
	\mathcal{P}_{{\rm T}} \simeq \frac{16 G_{\rm N} H^2 }{\pi}
	\left[1-2\left(1+C\right)\varepsilon_1\right].
	\label{eq4007}
\end{equation}

\subsection{Numerical calculations}

In this section, we present the results for the observational parameters and the number of e-folds, obtained by numerically solving the equation for the Hubble expansion rate, equation (\ref{hthetatheta}), for fixed $\eta$. This equation is a second-order differential equation with respect to the tachyon field $\theta$. To solve this equation numerically, we need to choose initial conditions, i.e., to fix the values of $h$ and $h_{,\theta}$ for a chosen value $\theta$. Let these be the values $h_{\rm i}\equiv h(\theta_{\rm i})$ and $h_{,\theta\rm i}\equiv h_{,\theta}(\theta_{\rm i})$ that correspond to the value of the field $\theta_{\rm i}$ at the time of the beginning of inflation ($\theta_{\rm i}=\theta(t_{\rm i})$). For this specific choice, we will fix the initial conditions  from the values of the observational parameters. As we will see, this approach allows us to estimate the range of parameter $\eta$ in the model with the corresponding predictions that should be in good agreement with observational data. For numerical solving of (\ref{hthetatheta}), bearing in mind form of (\ref{dottheta}), without loss of generality, we can set $t_{\rm i}=0$ and $\theta_{\rm i}=0$.

With the help of expressions (\ref{eq3007}) and (\ref{eq4007}), the observational parameters, the scalar spectral index $n_{\rm s}$, and the tensor-to-scalar ratio $r$ have been calculated in Ref.\ [\citen{Bertini:2020dvv}]
for the tachyon model in the holographic braneworld.
 Up to the second order in the slow-roll parameters, one finds
\begin{equation}
r= 16\varepsilon_{1{\rm i}}\left[1+C\varepsilon_{2{\rm i}}-\frac{2(2-h^2)}{3(4-h^2)}\varepsilon_{1{\rm i}}\right],
\label{r1}
\end{equation}
\begin{equation}
n_{\rm s}= 1-2\varepsilon_{1{\rm i}}-\varepsilon_{2{\rm i}}-\left(2-\frac{8h^2}{3(4-h^2)^2}\right)\varepsilon_{1{\rm i}}^2-\left(3+2C-\frac{2(2-h^2)}{3(4-h^2)}\right)\varepsilon_{1{\rm i}}\varepsilon_{2{\rm i}}-C\varepsilon_{2{\rm i}}\varepsilon_{3{\rm i}},
\label{ns1}
\end{equation}
where $\varepsilon_{i{\rm i}}\equiv\varepsilon_{i}(\theta_{\rm i})$ are the values of the slow-roll parameters at the beginning of inflation. The expressions (\ref{r1}) and (\ref{ns1}) at linear order in the slow-roll parameters $\varepsilon_{i}$ agree with the expressions obtained in the standard scalar field inflation with a tachyon field (see e.g. Ref.\ [\citen{Steer:2003yu}])
\begin{equation}
r\simeq 16\varepsilon_{1{\rm i}},\label{r}
\end{equation}
\begin{equation}
n_{\rm s}\simeq 1-2\varepsilon_{1{\rm i}}-\varepsilon_{2{\rm i}}.
\label{ns}
\end{equation}
The calculated parameters given in (\ref{r1}) and (\ref{ns1}) may be confronted with the observed values from
the Planck data [\citen{Planck:2018jri}]
\begin{equation}
n_{\rm s}=0.9649\pm 0.0042,
\end{equation}
\begin{equation}
r<0.056.
\end{equation}

Now, we calculate a functional relationship between the observational parameters, given by expressions at linear order in the slow-roll parameters, and the initial conditions $h_{\rm i}$ and $h_{,\theta\rm{i}}$. From (\ref{r}) and (\ref{ns}), using  (\ref{e22}), we obtain 
\begin{equation}
\eta=\frac{1}{2}\left(1-n_{\rm s}-\frac{r}{8}+\frac{h_{\rm i}^2}{(1-\frac{1}{2}h_{\rm i}^2)(1-\frac{1}{4}h_{\rm i}^2)}\frac{r}{32}\right).
\label{etain}
\end{equation}
The term proportional to $h_{\rm i}^2$,  that results from (\ref{e22}), 
 affects the value of the parameter $\eta$, which is another crucial difference in this model compared to the model in standard cosmology. Using (\ref{e1}), the expression (\ref{r}) can be rewritten as
\begin{equation}
h_{,\theta\rm i}=-\sqrt{\frac{3}{32}r\frac{1-\frac{1}{4}h_{\rm i}^2}{1-\frac{1}{2}h_{\rm i}^2}h_{\rm i}^4},
\label{hthetain}
\end{equation}
where we have chosen the minus sign, according to the demand that $\dot{h}<0$. Clearly, for each choice of $n_{\rm s}$, $r$ and $h_{\rm i}$ the parameter $\eta$ and the value $h_{,\theta\rm i}$ are fixed by (\ref{etain}) and (\ref{hthetain}). 

\subsection{Numerical results}

Now, we present the results for some fixed values of the parameters. In the following, we treat $h_{\rm i}$, besides $\eta$, as an additional free parameter of the model, the value of which lies in the interval $0<h_{\rm i}<\sqrt{2}$.  

\begin{figure}[h!]
\begin{center}
\includegraphics[scale=0.39]{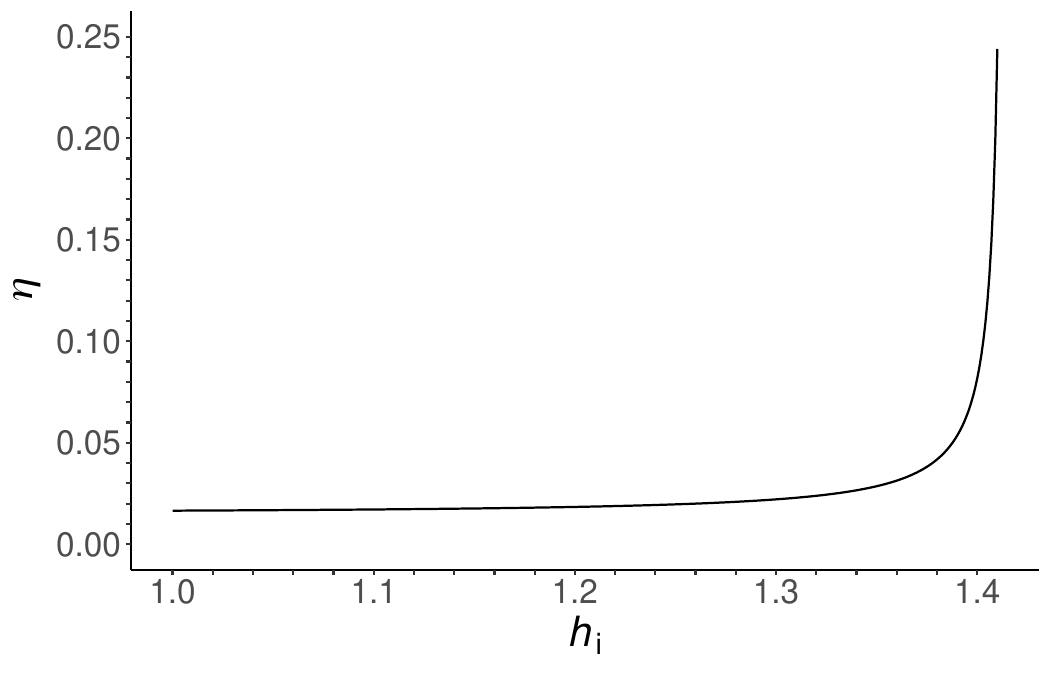}
\hspace{0.5cm}
\includegraphics[scale=0.39]{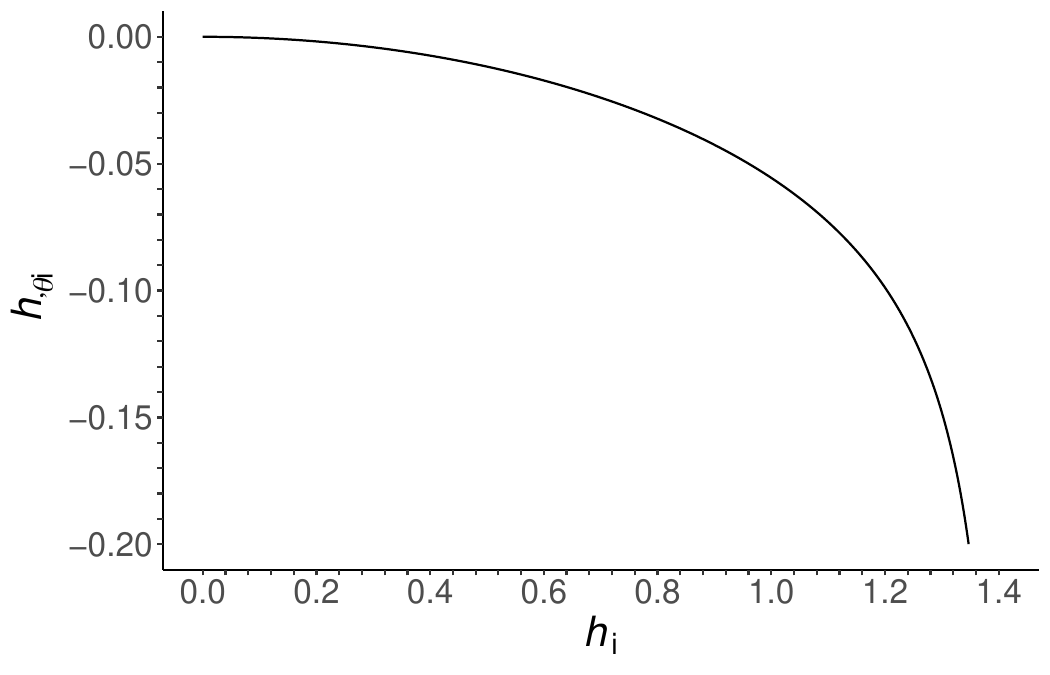}
\caption{$\eta$ versus $h_{\rm i}$ (left panel) and $h_{,\theta\rm i}$ versus $h_{\rm i}$ (right panel) for fixed $n_{\rm s}=0.966$ and $r=0.022$.}
\label{fig1}
\end{center}
\end{figure}

In Fig. \ref{fig1} we plot $\eta$ and $h_{,\theta\rm i}$ as function of $h_{\rm i}$ for fixed and arbitrary $n_{\rm s}=0.966$ and $r=0.022$. As we shall see, for the values of the observational parameters, which are well inside the observational region, the model can provide the inflation stage with $N(\theta_{\rm f})\approx 60$. We will use these values to illustrate further results.
It can be seen from Fig. \ref{fig1} that the value of parameter $\eta$ changes significantly only for those values of $h_{\rm i}$ which are near to the maximum value $h_{\rm i\;max}=\sqrt{2}$. For the other allowed values for $h_{\rm i}$, the parameter $\eta$ is almost constant and very close to the value obtained in the standard tachyon constant-roll inflation ($\eta\approx 0.0153$) [\citen{Mohammadi:2018oku}]. As shown in Fig. \ref{fig1}, $|h_{,\theta\rm i}|$ increases with $h_{\rm i}$ which indicates that there may be a deviation in the duration of inflation (e-folds number) in our model compared to the tachyon constant-roll model in standard cosmology [\citen{Mohammadi:2018oku}]. We will discuss those results in the next section. 

\begin{figure}[h!]
\begin{center}
\includegraphics[scale=0.39]{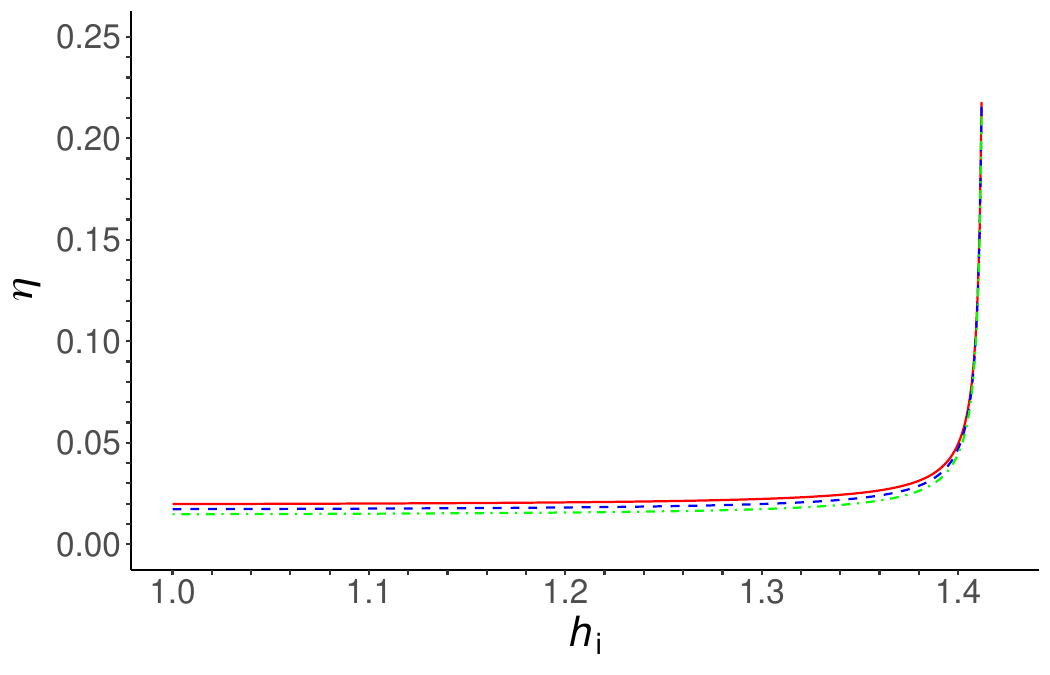}
\hspace{0.5cm}
\includegraphics[scale=0.39]{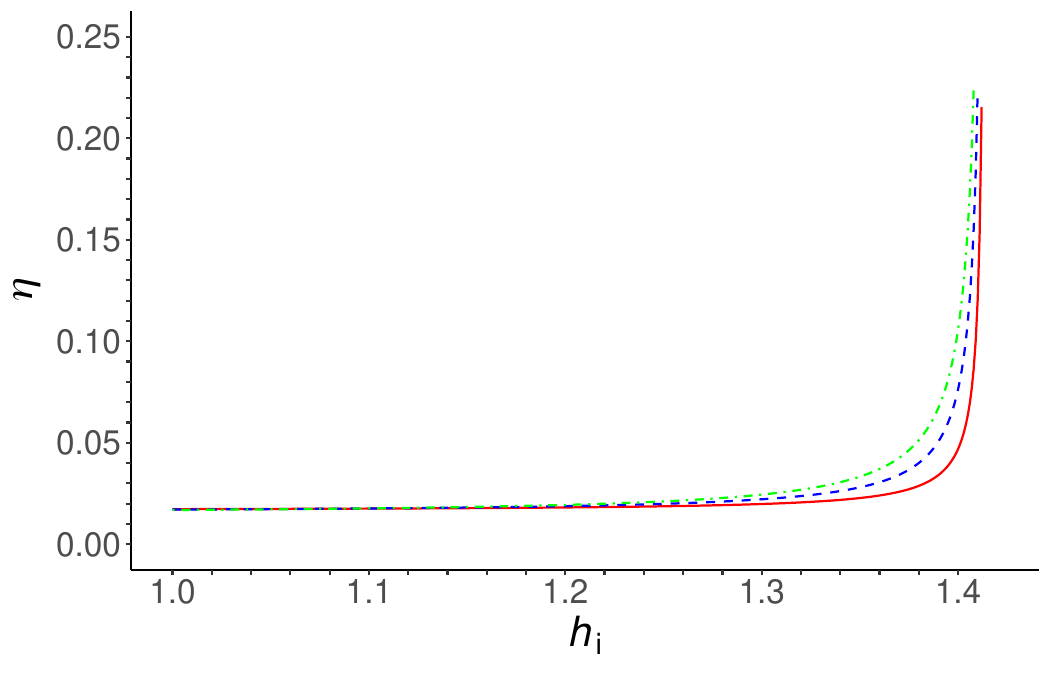}
\caption{$\eta$ versus $h_{\rm i}$. Left panel: $r= \textnormal{const}=0.01$, $n_{\rm s}=0.96$ (solid red line), $n_{\rm s}=0.965$ (dashed blue line), $n_{\rm s}=0.97$ (dash-dotted green line).
Right panel: $n_{\rm{s}}= \textnormal{const}=0.965$, $r=0.01$ (solid red line), $r=0.02$ (dashed blue line), $r=0.03$ (dash-dotted green line). }
\label{fig2}
\end{center}
\end{figure}

In Fig. \ref{fig2}, we plot $\eta$ as a function of $h_{\rm i}$ varying the values of the observational parameters. Again, the most significant range of the values of $h_{\rm i}$ is near its maximum, and the value of the $\eta$ is more affected by the change of $r$ than by the change of $n_{\rm s}$.

\begin{figure}[h!]
\begin{center}
\includegraphics[scale=0.39]{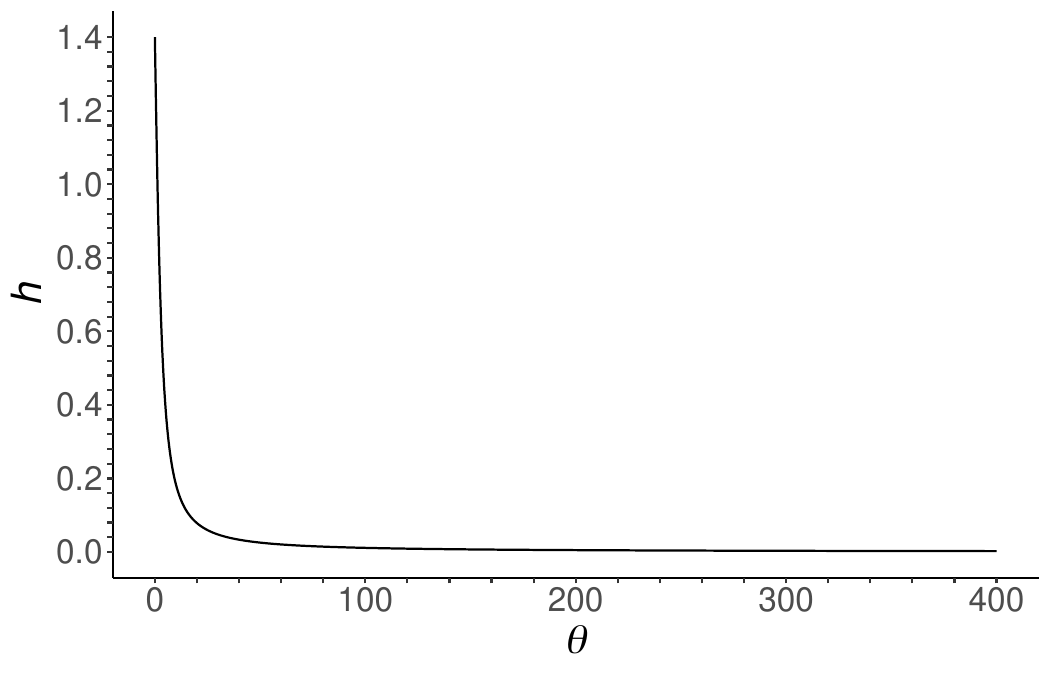}
\hspace{0.5cm}
\includegraphics[scale=0.39]{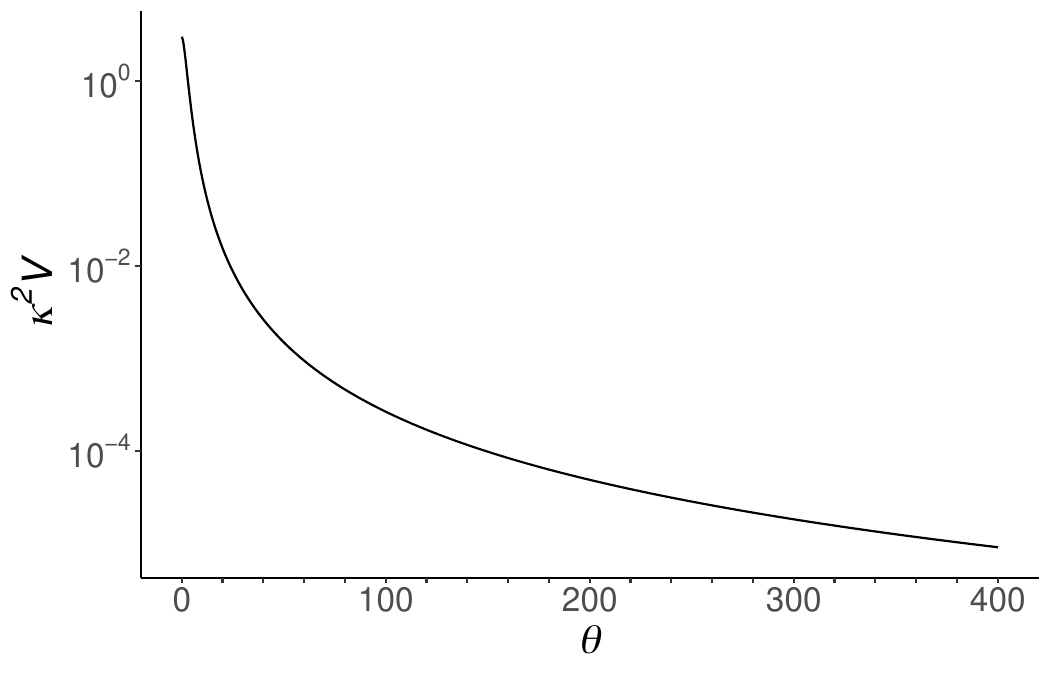}
\caption{The expansion rate (left panel) and the reconstructed potential (right panel) versus tachyon field in the model with $\eta=0.0817$, $h_{\rm i}=1.4$ and $h_{,\theta\rm i}=-0.4495$.}
\label{fig3}
\end{center}
\end{figure}

In Fig. \ref{fig3} we present a numerical solution to the equation (\ref{hthetatheta}) and the reconstructed potential, obtained for $n_{\rm s}=0.966$, $r=0.022$ and $h_{\rm i}=1.4$. For this choice of the parameters, using expressions (\ref{etain}) and (\ref{hthetain}), we find $\eta=0.0817$ and $h_{,\theta\rm i}=-0.4495$. It is obvious from Fig. \ref{fig3} that the reconstructed potential has all properties of a tachyon potential, defined by (\ref{vprop}). 

\begin{figure}[h!]
\begin{center}
\includegraphics[scale=0.39]{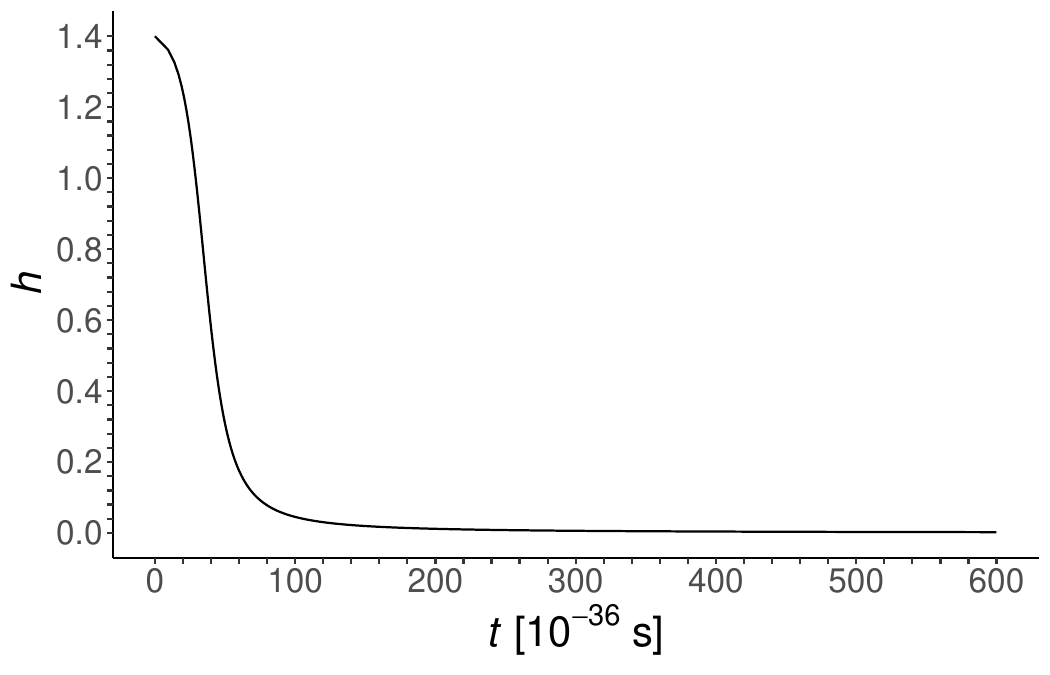}
\hspace{0.5cm}
\includegraphics[scale=0.39]{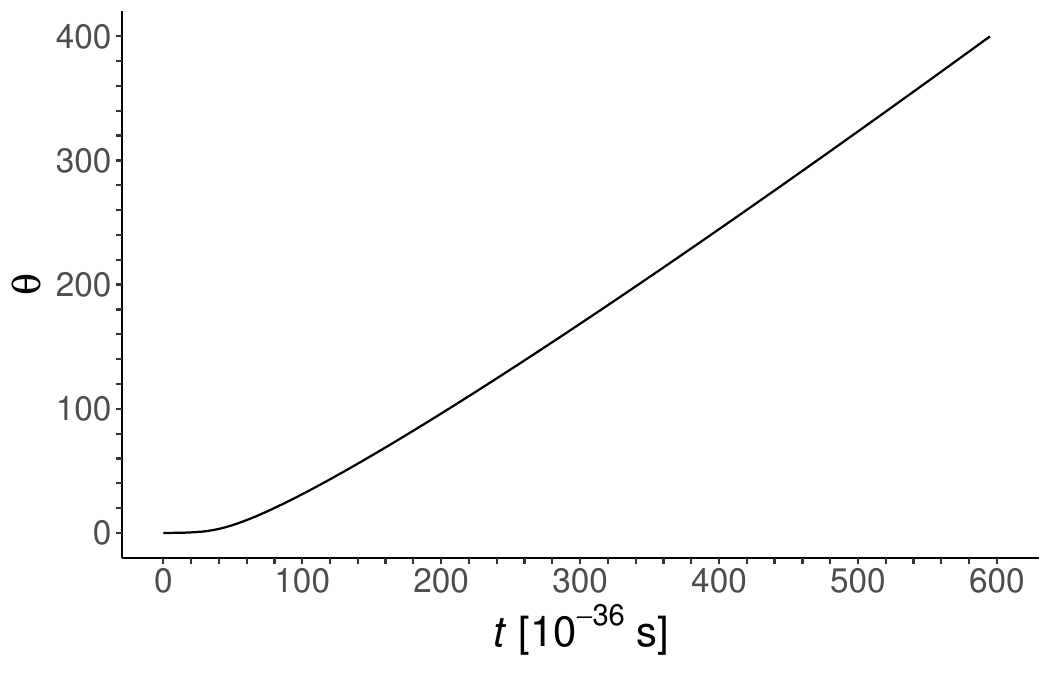}
\caption{The time evolution of the expansion rate (left panel) and the tachyon field (right panel) in the model with $\eta=0.0817$, $h_{\rm i}=1.4$ and $h_{,\theta\rm i}=-0.4495$.}
\label{fig4}
\end{center}
\end{figure}

The expression (\ref{dottheta}) may be integrated numerically, yielding the time $t$ as a function of $\theta$. With its help, the time evolution of the expansion rate and the tachyon field can be plotted (Fig. \ref{fig4}). 

\begin{figure}[h!]
\begin{center}
\includegraphics[scale=0.4]{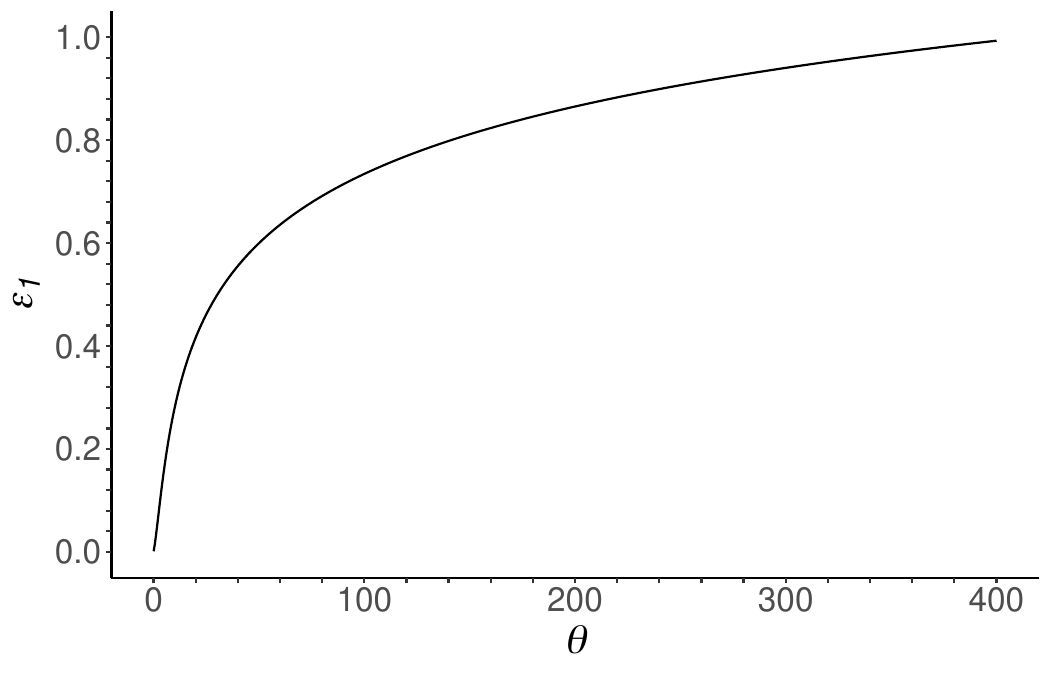}
\hspace{0.5cm}
\includegraphics[scale=0.4]{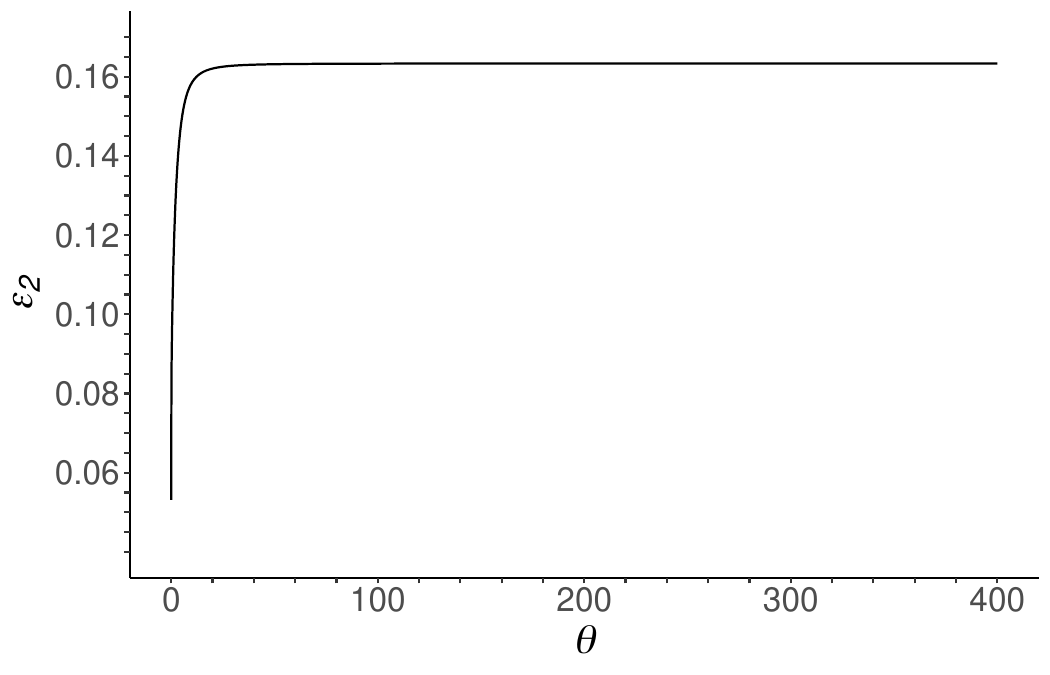}
\caption{The slow-roll parameters $\varepsilon_{1}$ and $\varepsilon_{2}$ versus tachyon field calculated numerically in the model with $\eta=0.0817$, $h_{\rm i}=1.4$ and $h_{,\theta\rm i}=-0.4495$.}
\label{fig5}
\end{center}
\end{figure}

The slow-roll parameters $\varepsilon_{1}$ and $\varepsilon_{2}$ as functions of $\theta$ are shown in Fig. \ref{fig5}. 

The parameter $\varepsilon_{1}$ increases and exceeds unity (for $\theta_{\rm f}=415.5$), defining the end of inflation. As already mentioned, for $h^{2}\ll 1$ the value $\varepsilon_{2}$ is almost constant. Furthermore, it is obvious from the shape of the curve in Fig. \ref{fig5} that the parameter $\varepsilon_{2}$ changes significantly only at the beginning of inflation.

Now, we calculate the number of e-folds associated with the observational parameters, the values of which we choose according to the restrictions from the observational data. 
Using (\ref{efolds}) we define $N\equiv N(\theta_{\rm f})$, 
where $\theta_{\rm f}\equiv\theta(t_{\rm f})$ is fixed by $\varepsilon_{1}(\theta_{\rm f})=1$. To calculate the number of e-folds, we proceed as follows. For some chosen  $n_{\rm s}$, $r$ and $h_{\rm \rm i}$, using (\ref{etain}) and (\ref{hthetain}), we calculate the parameter $\eta$ and the initial value $h_{,\theta\rm i}$. Then, we obtain $h(\theta)$ solving (\ref{hthetatheta}), for given $\eta$, with the initial conditions $h_{\rm i}$ and $h_{,\theta\rm i}$. From the condition for the end of inflation ($\varepsilon_{1}(\theta_{\rm f})=1$) we find $\theta_{\rm f}$. Finally, we use numerical integration in expression (\ref{efolds}) to find $N$. For $n_{\rm s}=0.966$, $r=0.022$ and $h_{\rm i}=1.4$ the number of e-folds is $N=60.2$.

\begin{figure}[h]
\begin{center}
\includegraphics[scale=0.36]{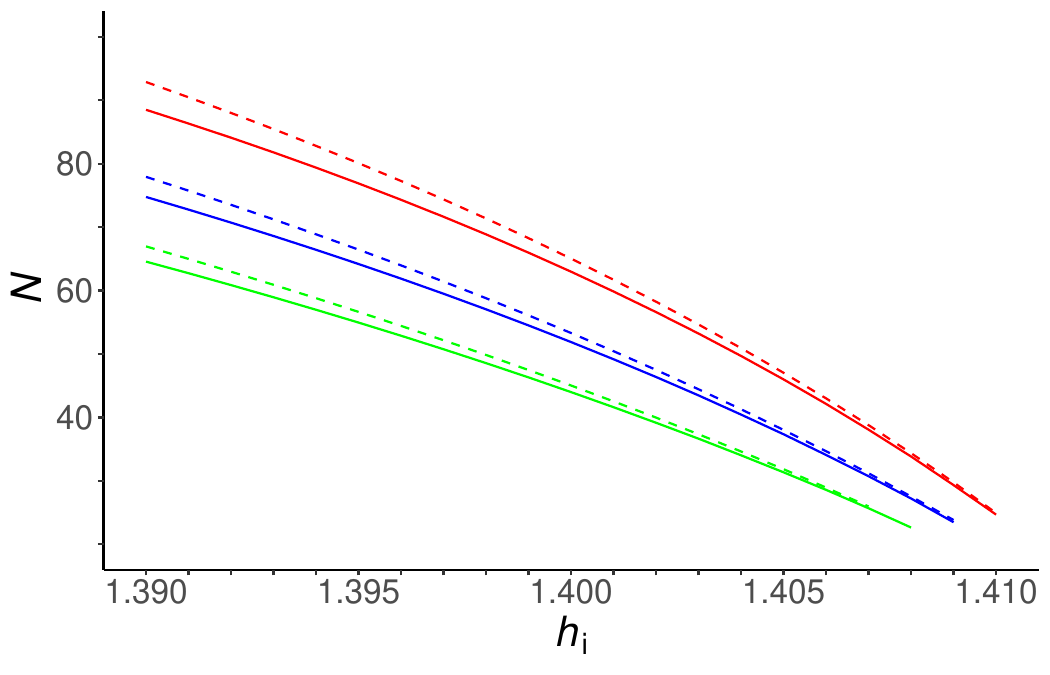}
\hspace{0.5cm}
\includegraphics[scale=0.36]{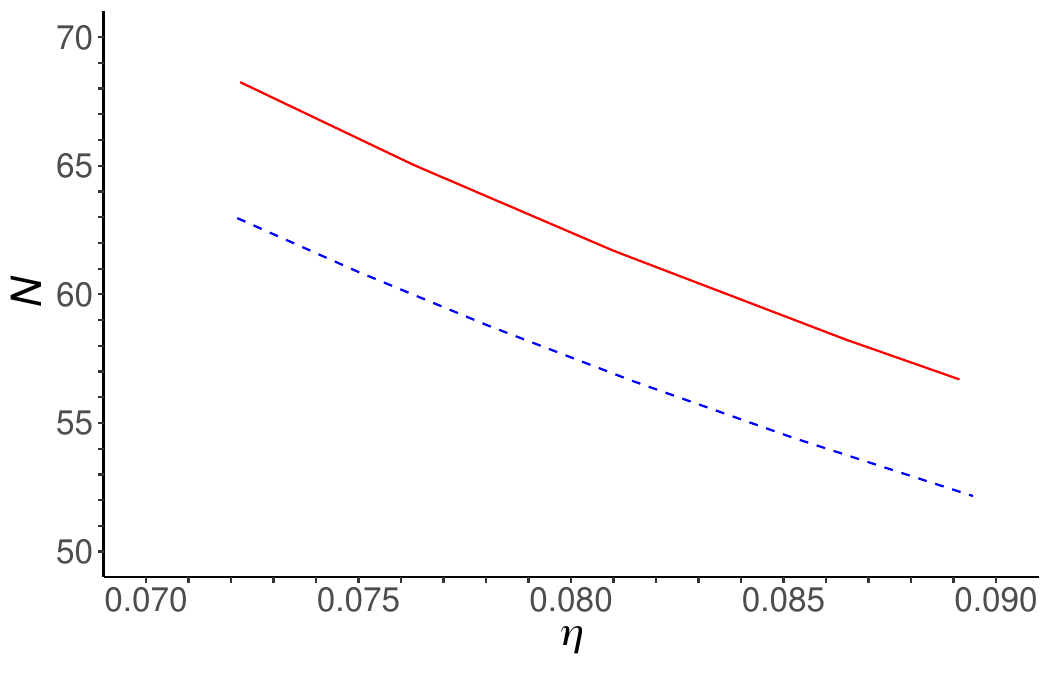}
\caption{Theoretical prediction for the functions $N$ of $h_{\rm i}$ (left panel)
and $N$ of $\eta$ (right panel) for several fixed observational parameters.
On the left panel,   $r=0.02$ (red line), $0.025$ (blue line), $0.03$ (green line), where  $n_{\rm s}=0.96$ is represented by solid line and  $n_{\rm s}=0.965$ by dashed line. The right panel shows the function $N$ of $\eta$  for randomly chosen observational parameters 
$n_{\rm s}=0.965$ and $r=0.02$ (solid red line) and $n_{\rm s}=0.965$ and $r=0.03$ (dashed blue line).}
\label{fig6}
\end{center}
\end{figure}

In Fig. \ref{fig6}, we plot $N$ as a function of $h_{\rm i}$ and as a function of $\eta$. Note that the change in the number $N$ due to the variation of the parameter $n_{\rm s}$ is significant only for small values of $h_{\rm i}$. This result is expected, considering how the parameter $\eta$ depends on the observable parameters $n_{\rm s}$ and $r$ (see Fig. \ref{fig2}). As mentioned earlier, the number of e-folds is also affected by the value of $h_{\rm i}$. Clearly, as the value of the parameter $h_{\rm i}$ increases, the number of e-folds decreases. Therefore, the interval of $h_{\rm i}$ in which the model gives inflation with $N\simeq 60$ is narrower than the model allows (see Fig. \ref{fig1}). In addition, Fig. \ref{fig6} indicates that the value of the parameter  $\eta$ should be close to $\eta\approx 0.08$ to have inflation with $N\simeq 60$.  
In comparison, the tachyon constant-roll inflation in standard cosmology for the values $n_{\rm s}=0.9615$ and $r=0.0197$, with $\eta=0.0153$, predicts inflation with $N=121$ [\citen{Mohammadi:2018oku}]. In our model, the chosen values of the observational parameters inside the allowed range of the Planck data, with $h_{\rm i}=1.4$ (and $\eta=0.077$), correspond to $N=64.4$.

\subsubsection{Estimate of the scale}
\label{estimate}
Here we estimate the radius of the AdS$_5$ curvature $\ell$ 
by making use of the normalization of the curvature perturbation spectrum.
We require that
the curvature perturbation spectrum $\mathcal{P}_{\rm S}$ at the 
CMB pivot scale $q_{\rm CMB}=0.05 {\rm Mpc}^{-1}$ coincides with the the power spectrum amplitude $A_s=2.10\pm 0.03\times 10^{-9}$
measured by Planck Collaboration 2018 [\citen{Planck:2018jri}].
For our purpose, we can assume $c_{\rm s}\simeq 1$ and approximate (\ref{eq3007}) by
\begin{equation}
	\mathcal{P}_{\rm S}(q)\simeq \frac{1}{\pi \varepsilon_1}
	\frac{H^2}{m_{\rm Pl}^2}=\frac{l_{\rm Pl}^2}{\pi \ell^2}
	\frac{h^2}{\varepsilon_1}	
	\label{eq5021}
\end{equation}
where for each $q$, the quantities $h(\theta)$ and $\varepsilon_1(\theta)$ take on their horizon crossing values at the corresponding $N(\theta)$. 
 In particular, we will use (\ref{eq5021}) at $q=q_{\rm CMB}$.
 
First, we find an approximate analytical function $N=N(\theta)$ using (\ref{efolds}) and an approximate solution to (\ref{hthetatheta}) (deduced from Fig.\ \ref{fig3})
\begin{equation}
	h\simeq h_{\rm i} e^{-\theta/(\alpha\ell)},
\end{equation}
where $\alpha$ is a constant. 
Then, from (\ref{efolds}), we find
\begin{equation}
	N(\theta)\simeq\frac{3\alpha^2 h_{\rm i}^2}{4}(1- e^{-2\theta/(\alpha\ell)}).
\end{equation}
Requiring $N(\theta_{\rm f}\simeq 60)$ at large $\theta_{\rm f}$ and taking $h_{\rm i}=1.4$, we find
$\alpha=6.4$, and hence 
\begin{equation}
h\simeq 1.4 e^{-\theta/(6.4\ell)}, \quad	N(\theta)\simeq 60(1- e^{-\theta/(3.2\ell)}).
\label{eq18}
\end{equation}

Next, for $q_{\rm CMB}$, we estimate the corresponding $N_{\rm CMB}\equiv N(\theta_{\rm CMB})$  at the horizon crossing, i.e., at the moment when
\begin{equation}
a_0 e^{N_{\rm CMB}}H_{\rm CMB}= q_{\rm CMB}
\label{eq19}
\end{equation}
where $H_{\rm CMB}\equiv H(\theta_{\rm CMB})$.
We assume that the initial conditions at $a=a_0$ are set in the deep subhorizon region where 
$q_{\rm CMB}\gg a_0 H_{\rm in}$.
Hence,   we set
\begin{equation}
	a_0  H_{\rm in}=\epsilon q_{\rm CMB} ,
	\label{eq20}
\end{equation}
where $\epsilon$ is a small parameter, e.g., of the order $\epsilon\sim 0.001$.
Combining Eqs.\ (\ref{eq19}) and (\ref{eq20}) we obtain 
\begin{equation}
	\epsilon e^{N_{\rm CMB}}\frac{H_{\rm CMB}}{H_{\rm in}} 
 =1.
	\label{eq21}
\end{equation}
Approximating $H_{\rm CMB}\simeq H_{\rm in}$, we obtain a rough estimate \begin{equation}
  N_{\rm CMB}\simeq -\ln \epsilon \simeq 7 	.
 	\label{eq22}
 \end{equation}
Then, from (\ref{eq18}) we find $\theta_{\rm CMB}\simeq 0.4\ell$ 
and $h_{\rm CMB}\simeq 1.3$. Finally, using (\ref{eq5021}) and equating 
$\mathcal{P} (q_{\rm CMB})= 2.1\times 10^{-9}$, we obtain an estimate
\begin{equation}
\ell \simeq 4 \times 10^5 l_{\rm Pl}.
\end{equation}

\subsection{Comparison with the Planck data}

\begin{figure}[h]
\begin{center}
\includegraphics[scale=0.75]{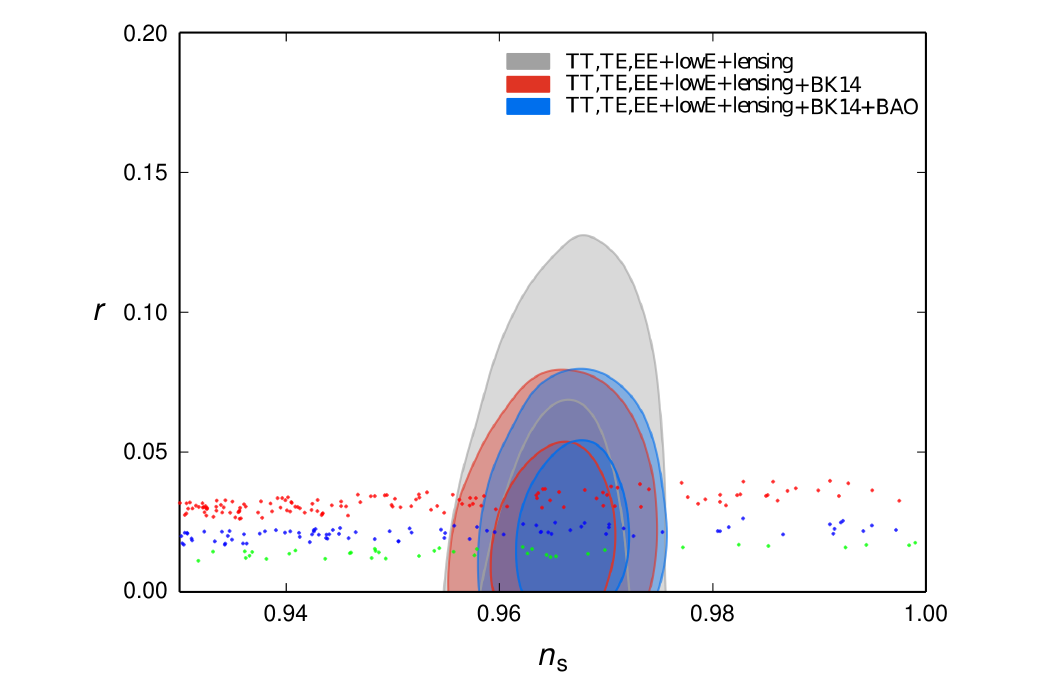}
\caption{$r$  versus $n_{\rm s}$ diagrams, in the models with $\eta={\rm const}$, with observational constraints from  Planck Collaboration 2018 [\citen{Planck:2018jri}]. The dots  are obtained for randomly chosen $\eta$, $h_{\rm i}$ and $h_{,\theta\rm i}$ in the intervals $0.08\leq\eta\leq 0.085$, $1.39\leq h_{\rm i}\leq 1.41$ and $-0.01\leq h_{,\theta\rm i}\leq-0.9$. The colors represent different $N$: $N=55$ (red), $N=60$ (blue), and $N=65$ (green).}
\label{fig7}
\end{center}
\end{figure}

Note that previous results for the number of e-folds were obtained using the functional relationship between the observational parameters and the initial conditions $h_{\rm i}$ and $h_{,\theta\rm{i}}$, equations (\ref{etain}) and (\ref{hthetain}). More accurate results can be obtained using (\ref{r1}) and (\ref{ns1}) and applying the standard approach based on calculating observation parameters $n_{\rm s}$ and $r$ for a given e-folds number $N$, described in details in refs. [\citen{Bilic:2018uqx}], [\citen{Milosevic:2018gck}] and [\citen{Bilic:2016fgp}]. For randomly chosen $\eta$, $h_{\rm i}$, and $h_{,\theta\rm i}$,  the equation (\ref{hthetatheta})  and the differential equation for the number of e-folds $dN/d\theta=h/(\ell\dot{\theta})$, are solved simultaneously from $\theta_{\rm i}=0$ to some arbitrary value $\theta_{\rm f}$ which is large enough, to provide the end of inflation ($\varepsilon_{1}(\theta_{\rm f})=1$). Then, demanding $N=N(\theta_{\rm i})-N(\theta_{\rm f})$ the value $\theta_{\rm i}$ is obtained for given $N$. The value  $\theta_{\rm i}$ is then used to find $\varepsilon_{i\rm{i}}$ and observational parameters (Fig. \ref{fig7}). According to the previous considerations, we choose the values for the free parameters in ranges $0.08\leq\eta\leq 0.085$ and $1.39\leq h_{\rm i}\leq 1.41$. Initial values of $h_{,\theta\rm{i}}$, which must be negative, are assumed to be in the interval $-0.01\leq h_{,\theta\rm i}\leq-0.9$ provided that $\varepsilon_{1}(\theta_{\rm{i}})\ll 1$. 
\begin{figure}[h!]
\begin{center}
\includegraphics[scale=0.75]{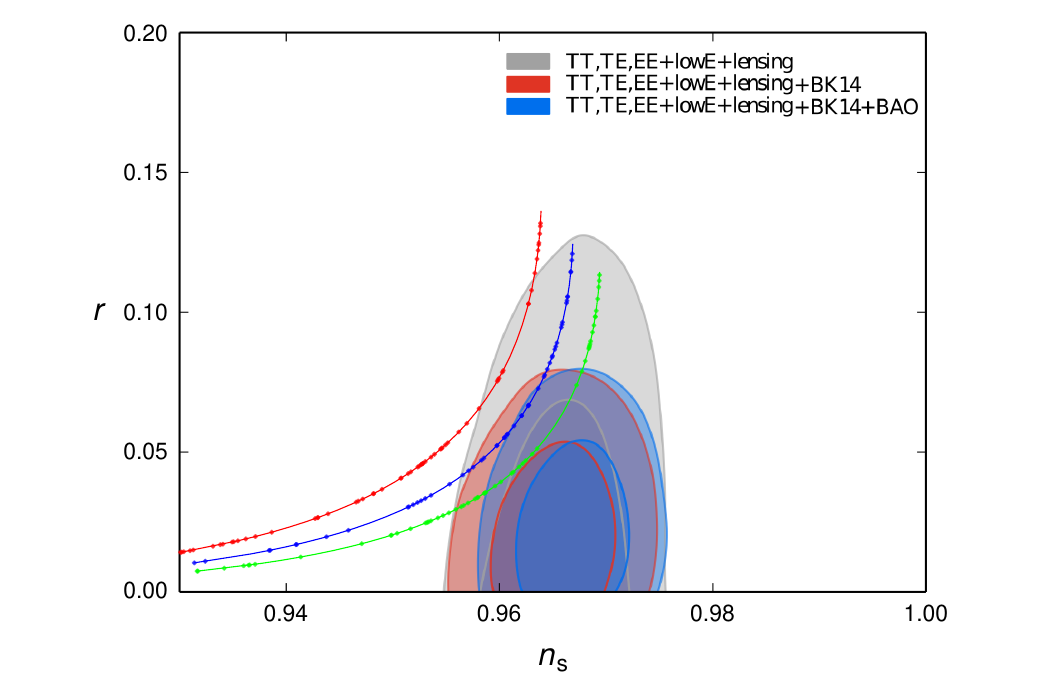}
\caption{$r$  versus $n_{\rm s}$ diagrams, in the models with $\bar{\eta}={\rm const}$, with observational constraints from  Planck Collaboration 2018 [\citen{Planck:2018jri}]. The dots  are obtained for randomly chosen $\bar{\eta}$, $h_{\rm i}$ and $h_{,\theta\rm i}$ in the intervals $-0.001\leq\bar{\eta}\leq -0.04$, $1.38\leq h_{\rm i}\leq1.414$ and $-0.1\leq h_{,\theta\rm i}\leq-0.9$. The colors represent different $N$: $N=55$ (red), $N=60$ (blue), and $N=65$ (green). The analytical results obtained in Ref.\ [\citen{Stojanovic:2023qgm}] are depicted by the full lines.}
\label{fig8}
\end{center}
\end{figure}
We can easily extend our consideration to other constant-roll models using the above procedure for calculating the observational parameters. The second slow-roll parameter can be defined in terms of the Hubble expansion rate only
\begin{equation}
\bar{\eta}=-\frac{\ell\ddot{h}}{2h\dot{h}}.
\label{defbareta}
\end{equation}
The parameter $\bar{\eta}$ is different from $\eta$ except in the canonical single field inflation where these two parameters are equivalent [\citen{Stojanovic:2023qgm}]. Using the same procedure as in Section 3, we derive a differential equation for the Hubble expansion rate $\bar{\eta}$
\begin{equation}
hh_{,\theta\theta}-h,_{\theta}^2\left(1+\frac{h^2}{4(1-\frac{1}{2}h^2)(1-\frac{1}{4}h^2)}\right)-\frac{3}{2\ell^2}\bar{\eta} h^4\frac{1-\frac{1}{4}h^2}{1-\frac{1}{2}h^2}=0.
\end{equation}
By applying the numerical procedure, we calculate the observational parameters from (\ref{r1}) and (\ref{ns1}) in the model with $\bar{\eta}=\rm{const}$, where $\varepsilon_{2}$ and $\varepsilon_{3}$ are given by
\begin{equation}
\varepsilon_{2}=2(\varepsilon_{1}-\bar{\eta}),
\end{equation}
\begin{equation}
\varepsilon_{3}=2\varepsilon_{1}.
\end{equation}
As expected, the prediction for the observational parameters (Fig. \ref{fig8}) in the holographic braneworld agrees with the model-independent prediction in Ref.\  [\citen{Stojanovic:2023qgm}], obtained analytically for any model that satisfies the condition $\bar{\eta}=\rm{const}$. As we will shortly see, this fact will be of crucial importance for consideration of the connection between our model and the swampland criteria in quantum gravity.

Unfortunately, the tachyon inflation suffers from a reheating problem
imminent for all tachyon models with the ground state at 
$\theta\rightarrow\infty$ [\citen{kofman}]. 
After the inflationary epoch, the tachyon will remain a dominant component  
unless at the end of inflation, it
decayed into inhomogeneous fluctuations and
other particles. This period, known as reheating
[\citen{kofman2}],  links
the inflationary epoch with the subsequent thermalized
radiation era. 
Unlike in the canonical models of inflation, the tachyon field rolls towards its ground state  
without oscillating about it, and the conventional reheating mechanism does not work.
However, it was shown [\citen{cline}] that coupling of massless fields to the time-dependent 
tachyon condensate could yield a reheating efficient enough to overcome the problem of a cold dark-matter dominance. 

The authors of Ref.\ [\citen{djordjevic1}]  have explicitly studied the reheating that results
from a coupling of the tachyon with a U(1) gauge field through a mechanism of cosmological production. 
Unfortunately, it turns out that the proposed scenario alone cannot solve the reheating problem of the tachyon inflation.
It would be of considerable interest to investigate the effects of
cosmological creation in the warm inflation models [\citen{berera}].
In warm inflation, radiation due to dissipative effects
is produced in parallel with the inflationary
expansion, and inflation ends when the universe heats up to become radiation-dominated.
This scenario has been successfully applied to tachyon inflation models
[\citen{herrera}] and, in principle, should also work for
tachyon inflation in
the holographic braneworld inflation presented here. However, this study is
beyond the scope of the present paper.

\section{The swampland criteria and the braneworld inflation}

It is known that the idea of swampland [\citen{Vafa:2005ui}] appeared in the context of the investigation of low-energy effective field theories (non)compatible with quantum gravity. Much attention and effort have been paid to the investigation of criteria [\citen{Obied:2018sgi}] related to "swampland" or to "string theory landscape" to distinguish between effective field theories that can or cannot  be UV-completed to a quantum theory of gravity.

Investigation of swampland criteria in cosmology, more precisely the theory of inflation, is pretty recent and presented in  very few research articles, especially in the case of braneworld and holographic inflation.
While some inflationary models are not compatible with these criteria, inflationary models in the braneworld scenario, in principle, have the potential to evade the swampland constraints [\citen{Mohammadi:2020ftb}]. 

Let us initiate consideration of the swampland criteria in the context of our RSII holographic constant-roll model. The swampland criteria read as follows [\citen{Kehagias:2018uem}]
\begin{equation}
\frac{|\Delta\theta|}{M_{\rm Pl}}<c,
\label{sc}
\end{equation}

\begin{equation}
M_{\rm Pl}\frac{|V_{,\theta}|}{V}>c',
\label{sc2}
\end{equation}
where $c$ and $c'$ are constants of order one and $M_{\rm Pl}=(8\pi G_{\rm N})^{-1/2}\simeq 2.4\cdot  10^{18}\;\rm{GeV}$. Further investigations have shown that the constant $c'$ could be of order  0.1 [\citen{Mohammadi:2020vgs}]. From the swampland criteria, it may be seen that the range traversed by the field is bounded from above, and there is a lower bound on the gradient of the potential. As pointed out in Refs. [\citen{Oikonomou:2023bmn}] and [\citen{Odintsov:2023aaw}], the swampland conjectures will be met if at least one of the conditions is satisfied. In the following, we will examine whether the second criterion (\ref{sc2}) is satisfied in our model. To the best of our knowledge, there has been no attempt in the literature to check the swampland criteria in a holographic inflation model with dynamics similar to the one in our model. 

Consider the second criterion.
In the single-field slow-roll inflation model the first slow-roll parameter is given by
\begin{equation}
	\varepsilon_{1}=\frac{M_{\rm Pl}^2}{2}\frac{V_{,\theta}^2}{V^2},
	\label{defe1V}
\end{equation}
and the second swampland criterion implays $\varepsilon_{1}>c'^{2}/2$ [\citen{Kinney:2018nny}]. Note that for $c'\sim 0.1$, one obtains a consistency of the model with both the observational data and the swampland conjecture.  
Namely,  as a consequence of (\ref{sc2}) and (\ref{defe1V}), to satisfy the second swampland criteria one needs $\varepsilon_{1}>0.005$,  in perfect agreement with the condition $\varepsilon_{1}<1$ for accelerated expansion.

Next, we demonstrate that in our holographic approach the parameter $\eta=\rm{const}$ is not consistent with the second swampland criterion, whereas $\bar{\eta}=\rm{const}$ may satisfies both the second swampland criterion and Planck data. The definition (\ref{defe1V}) is only valid in the slow-roll approximation, in a model with a canonical scalar field in the standard cosmology [\citen{Gron:2018rtj}]. For tachyon field the therm $V_{,\theta}/V$ can be calculated using the equation for $\theta(t)$ [\citen{Steer:2003yu}]
\begin{equation}
\frac{\ell\ddot{\theta}}{1-\dot{\theta}^2}+3h\dot{\theta}+\frac{\ell V_{,\theta}}{V}=0,
\label{eqtheta}
\end{equation}
yielding
\begin{equation}
\frac{\ell V_{,\theta}}{V}=-3h\dot{\theta}\left(1+\frac{\eta}{3(1-\dot{\theta}^2)}\right).
\label{ddota}
\end{equation}
For the small and constant $\eta$ it leads to
\begin{equation}
\frac{\ell V_{,\theta}}{V}\simeq-3h\dot{\theta}. 
\label{ddotaapp}
\end{equation}
By combining the time derivative of $h=\ell\dot{a}/a$ with Eq. (\ref{F2}), we obtain 
\begin{equation}
\frac{\ddot{a}}{a}=\frac{h^2}{\ell^2}\left(1-\frac{3}{2}\dot{\theta}^2\frac{1-\frac{1}{2}h^2}{1-\frac{1}{4}h^2}\right).
\label{ddota2}
\end{equation}
From this, it follows that the condition for accelerated expansion $\ddot{a}>0$
will be met if
\begin{equation}
	\dot{\theta}^2<\frac{2}{3}\frac{1-\frac{1}{4}h^2}{1-\frac{1}{2}h^2},
\end{equation}
and, when $h^{2}\ll 1$, inflation ends  for $\dot{\theta_{\rm f}}\simeq \sqrt{2/3}$. According to (\ref{ddotaapp}), the second swampland criteria requires $h_{\rm f}>0.1/\sqrt{6}\approx 0.04$. As we can see (Fig. \ref{fig3}), the calculated Hubble expansion rate $h$ tends to a very small value towards the end of inflation. One may conclude that the second criterion will not be satisfied for those values of $h$ that provide agreement  with Planck data. 

\begin{figure}[h!]
\begin{center}
\includegraphics[scale=0.39]{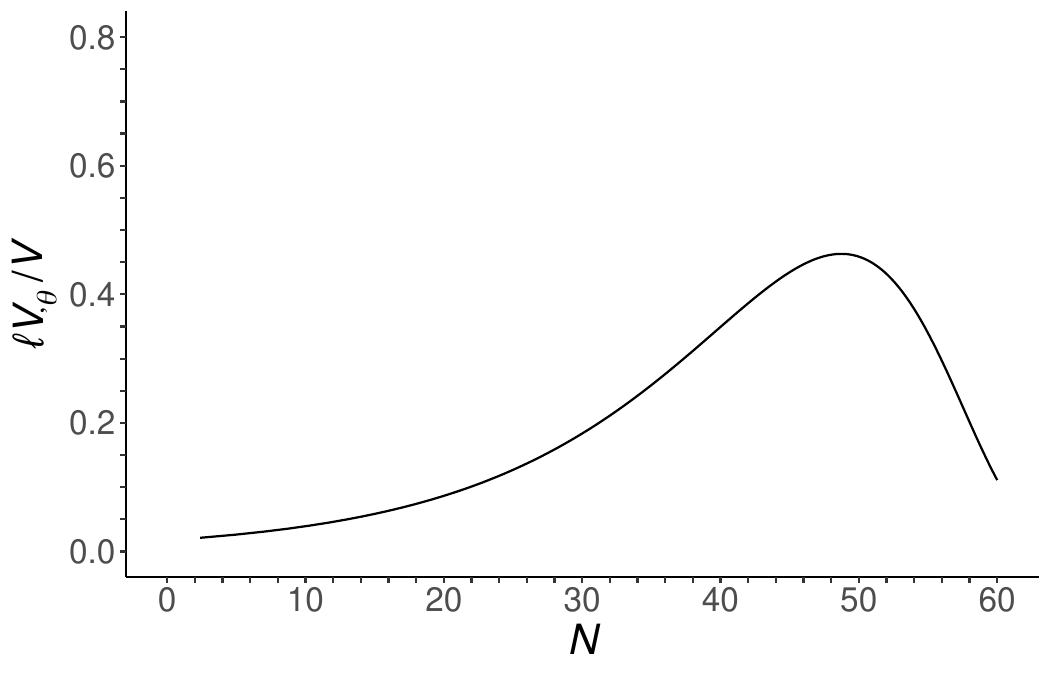}
\includegraphics[scale=0.39]{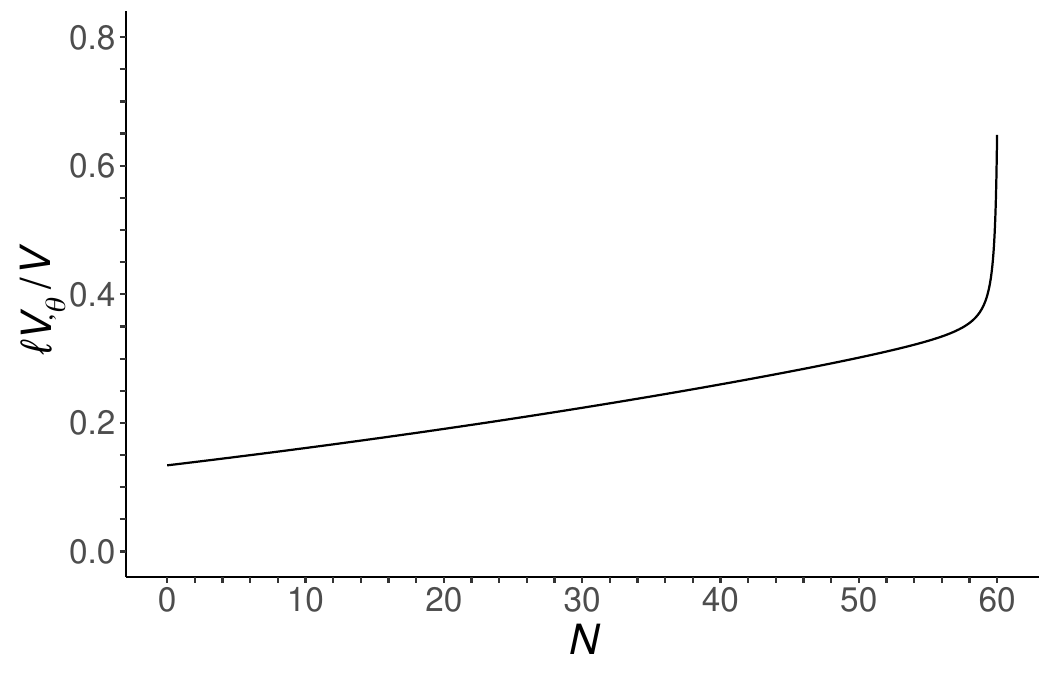}
\caption{The value of $\ell V_{,\theta}/V$ versus $N$ in the model with $\eta=\rm{const}$ (left panel)  and $\bar{\eta}=\rm{const}$ (right panel). The parameter values  are $\eta=0.0816$, $h_{\rm i}=1.4046$, $h_{,\theta\rm{i}}=-0.4503$ (left panel) and $\bar{\eta}=-0.0082$, $h_{\rm i}=1.2170$, $h_{,\theta\rm{i}}=-0.1964$ (right panel).}
\label{fig9}
\end{center}
\end{figure}

Next, consider the second swampland criterion using the parameter $\bar{\eta}$ given by (\ref{defbareta}), which does not depend explicitly on the field $\theta$! Remember that $\eta$ and $\bar{\eta}$ are the same only in canonical single-field inflation, while in our model, they are different, and, as we will see, it will lead us to a different scenario. In this case, using the derivative of (\ref{potential}) with respect to the tachyon field, we obtain 
\begin{equation}
	\frac{\ell V_{,\theta}}{V}=-\frac
	{(1-\frac{h^2}{2})\left(-36 h^3 (1-\frac{h^2}{4})-8\ell^{2}hh_{,\theta}^2+8\ell^{2}(1-\frac{h^2}{2})h_{,\theta\theta}\right)\ell h_{,\theta}}
	{18h^4(1-\frac{h^2}{4})^2-8\ell^{2}(1-\frac{h^2}{2})^2h_{,\theta}^2}.
\end{equation}
From Fig.\ \ref{fig9}, we can conclude that our holographic model provides the second swampland criterion to be satisfied during \textbf{whole} inflation for a wide range of the relevant parameters. For instance, for $\bar{\eta}=\rm{const.}$ and $N=65$, the following set of observational parameters is obtained: $n_{\rm s}=0.964$, $r=0.055$, that are in a good agreement with the current observational data. We can roughly estimate that the second swampland criterion is satisfied for $n_{\rm s}>0.96$ and $r>0.05$. A similar behavior for $V_{,\theta}/V$  is obtained in Ref.\  [\citen{Mohammadi:2020ftb}].  

\section{Summary and conclusions}

We have studied constant-roll inflation driven by the tachyon-field inflaton in the the holographic branewo\-rld. In particular, we have focused on two models with constant roll conditions $\eta=\rm{const}$ and $\bar{\eta}=\rm{const}$, respectively.
As expected, the inflation dynamics is modified due to the differences in the holographic Friedmann equations compared with those in the standard approach. Owing to a complicated form of the equation for the Hubble rate, we have mainly used numerical calculations throughout this work.  

We have compared the theoretical results for observational parameters with the Planck data. We have deduced optimal values of our free parameters for which the model predictions are in good agreement with the observational data. In comparison with the results of the tachyon constant-roll inflation in standard cosmology [\citen{Mohammadi:2018oku}], our model gives significantly lower values for the number of e-folds, much closer to a widely accepted $N\simeq 60$. 

We have found that the formulation in which the second slow-roll parameter $\bar{\eta}$ is constant (\ref{defbareta}) meets the swampland criteria. In this case, the obtained values for the observational parameters $n_{\rm s}$ and $r$ are still in the range of the Planck data but slightly less favorable than the values obtained with constant second slow-roll parameter $\eta$ (\ref{defetanew}). 

It is worth noting that the swampland "problem" for holographic inflation may be closely related to another fundamental problem of braneworld inflation. Namely, according to our model, inflation on the holographic brane occurs under the influence of the scalar inflaton field defined on the brane. However, after the Big Bang, in the era of quantum gravity, inflation might be driven by another field, for instance, some moduli scalar field in the bulk, as a possible manifestation of supergravity or superstring theory. In that case, an effective low-energy theory consistent with the swampland conjectures may not match completely with our effective gravity on the brane. 

As a point for more detailed consideration of the swampland criteria in brenworld inflation, note that the equations of the gravity field on the brane that lead to holographic cosmology are not derived strictly from some fundamental action analogous to the Einstein-Hilbert action. Given this point, the investigation of fulfilling the swampland criteria for the constant-roll inflation presented in Ref.\ [\citen{Stojanovic:2023qgm}] is a subject of ongoing research.

\section*{Acknowledgments}

This work has been supported by the ICTP-SEENET-MTP project NT-03 Cosmology-Classical and Quantum Challenges and the COST Action CA18108 "Quantum gravity phenomenology in the multi-messenger approach". M. Stojanovic acknowledges the support provided by The Ministry of Science, Technological Development and Innovation of the Republic of Serbia under contract 451-03-47/2023-01/2000113. D. D. Dimitrijevic, G. S. Djordjevic, and M. Milosevic acknowledge the support provided by the same Ministry 451-03-47/2023-01/2000124. In addition, G. S. Djordjevic acknowledges the support of the CEEPUS Program RS-1514-03-2223 "Gravitation and Cosmology" as well as the hospitalities of colleagues at the University of Banja Luka and CERN-TH.
N. Bili\' c acknowledges the hospitality of the Department of Physics at 
the University of Ni\v s, where part of his work was completed. 


\end{document}